\documentclass[9pt,aip,twocolumn,twoside, reprint,nofootinbib]{revtex4-1}

\usepackage{graphicx}
\usepackage{bm}
\usepackage{braket}
\usepackage{float}

\begin{document}

\preprint{AIP/123-QED}

\title{Two-Dimensional Photonic Crystals for Engineering Atom-Light Interactions}

\author{Su-Peng Yu*}
\affiliation{Norman Bridge Laboratory of Physics MC12-33, California Institute of Technology\\ Pasadena, California 91125, USA}

\author{Juan A. Muniz*}%
\affiliation{Norman Bridge Laboratory of Physics MC12-33, California Institute of Technology\\ Pasadena, California 91125, USA}

\author{Chen-Lung Hung}
\affiliation{Department of Physics and Astronomy, Purdue University,\\ West Lafayette, Indiana 47907, USA}
\affiliation{Purdue Quantum Center, Purdue University,\\ West Lafayette, Indiana 47907, USA}

\author{H. J. Kimble}%
\affiliation{Norman Bridge Laboratory of Physics MC12-33, California Institute of Technology\\ Pasadena, California 91125, USA}


\date{\today}

\begin{abstract}
We present a two-dimensional (2D) photonic crystal system for interacting with cold cesium (Cs) atoms. The band structures of the 2D photonic crystals are predicted to produce unconventional atom-light interaction behaviors, including anisotropic emission, suppressed spontaneous decay and photon mediated atom-atom interactions controlled by the position of the atomic array relative to the photonic crystal. An optical conveyor technique is presented for continuously loading atoms into the desired trapping positions with optimal coupling to the photonic crystal. The device configuration also enables application of optical tweezers for controlled placement of atoms. Devices can be fabricated reliably from a 200nm silicon nitride device layer using a lithography-based process, producing predicted optical properties in transmission and reflection measurements. These 2D photonic crystal devices can be readily deployed to experiments for many-body physics with neutral atoms, and engineering of exotic quantum matter.
\end{abstract}

\footnote{S.-P. Yu and J. A. Muniz contributed equally to this work}

\maketitle

\begin{quotation}
Specialized two-dimensional photonic crystals have been developed to interact with ultra-cold atoms, which are identical particles demonstrating quantum behavior both in their interaction with photons and in their motional degrees of freedoms. In the system presented here, the quantum nature of atoms is complemented with capabilities of 2D photonic crystals to engineer optical dispersion, light emission patterns, and photon-mediated coherent interactions. The combined system enables atom-atom interactions mediated by photons in the guided modes of the photonic crystals to provide new tools to engineer quantum many-body systems and create exotic quantum matters.
\end{quotation}

The introduction of nano-photonics to the field of quantum optics and atomic physics greatly broadens the capabilities of atom-photon systems \cite{Chang2018}. Interaction with photonic structures such as tapered optical fibers \citep{Goban2012,Sague_2007,Vetsch2010,Corzo2016}, micro-cavities \citep{Thompson2013,Alton2010,Volz2014}, and photonic crystal (PhC) waveguides \citep{Goban2014,Goban2015,Hood2016} enable engineering of atom optical properties by modifying the local density of state (LDOS) of the electromagnetic field that interacts with the atoms. Such engineering capabilities have been actively explored in various solid-state systems such as quantum dots \citep{Englund2008, Lodahl2014, Khitrova2007},  color centers \citep{Patel2016,Sipahigil2016}, and embedded rare-earth ions \citep{Riedmatten2015,Zhong2015,Dibos2017}. Among the variety of quantum emitters available nowadays, the identical-particle nature of neutral atoms provides particular advantages in forming quantum many-body systems. Both the center-of-mass motion of the atoms and the photons they interact with can potentially become components of such quantum systems \citep{Simon2011,Uehlinger2013}, and interaction between neutral atoms can be introduced by coupling them to shared photonic modes \citep{Tudela2015,Kollar2017}. Recently, there is an interest on exploring how 2D arrays of cold atoms coupled to a photonic bath show interesting topological properties robust under scattering \cite{Perczel_PRL2017,Shi2017}. The array of capabilities of atom-optics systems provide building blocks for exotic quantum many-body systems.

The cold atom community has put significant efforts into creating controllable interacting 2D systems \citep{Bloch2005}. Advances in optical lattice systems allow for creation of various geometries of lattices \citep{Becker2010,Mazurenko2017}, engineering of many-body Hamiltonians \citep{Jaksch2005}, and manipulating atoms on a single site \citep{Weitenberg2011}. The 2D platforms demonstrate physics that is not manifest in 1D geometries, such as frustrated spin systems \citep{Struck2011} and directional emission \citep{Tudela_PRL2017,Tudela_PRA2017}.

It is interesting to explore beyond the photonic cavity and 1D waveguide systems, and create equally versatile 2D photonic systems. The particular platform of interest here is the 2D photonic crystal slab, where a periodic spatial modulation in dielectric distribution is arranged over a dielectric slab. The properties of a photonic crystal can be exploited to provide sub-wavelength scale optical trapping and dispersion engineering \citep{Tudela2015,MunizPhD,YuPhD}. In this article, we present two distinct types of 2D photonic crystal slabs demonstrating novel photonic properties, such as directional spontaneous emission, spontaneous emission suppression, engineering of coherent atom-atom interactions, and novel topological properties in linear dielectric systems. Optical trapping schemes for trapping neutral ultra-cold Cs atoms in the vicinity of the photonic crystal slabs are also devised.

\begin{figure}[htb!]
\centering
\includegraphics[width=1\linewidth]{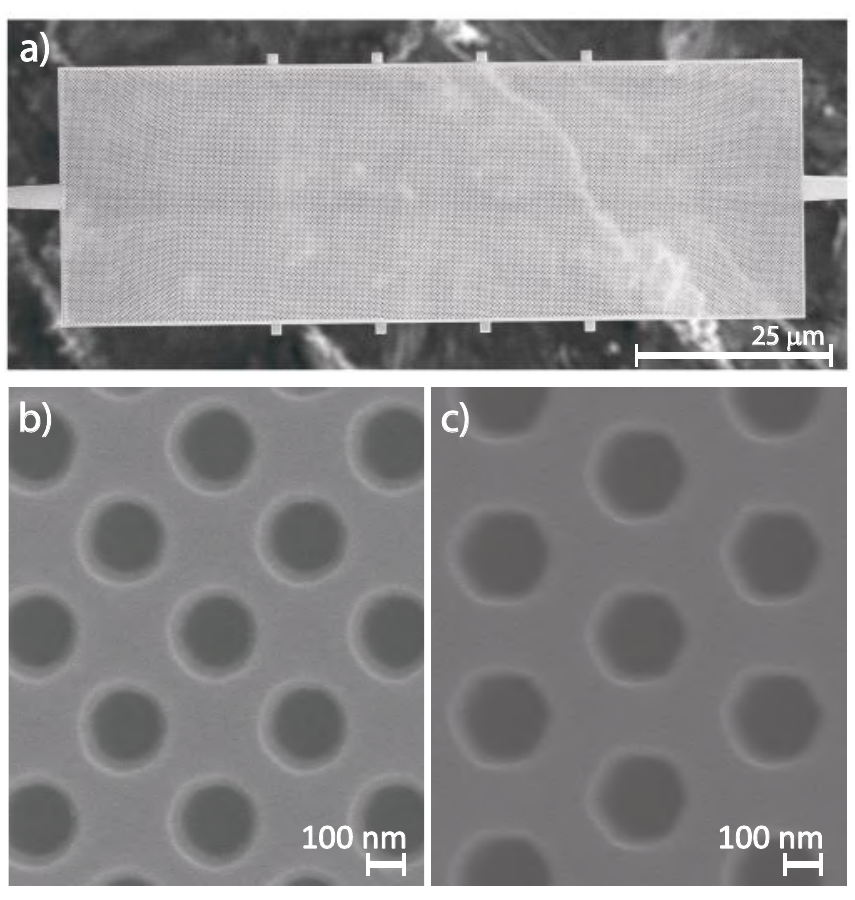}%
\caption{A suspended 2D PhC slab. (a) SEM image of a photonic crystal slab structure, suspended by two single-beam SiN waveguides on left and right edges. The slab contains multiple sections with smoothly varying crystal parameters for guiding purposes (see Fig.~\ref{fig:SimVsMeas} and discussions). The dielectric tabs spaced along the top/bottom of the slabs delineate the boundaries between such sections. Irregular pattern in the background is an aluminum stage for the SEM, seen through a through-window of the chip. (b) Zoom-in SEM image of the square lattice of circular holes and (c) triangular lattice of hexagonal holes photonic crystals.}
\label{fig:TDCexample}
\end{figure}

The 2D photonic crystal slabs are fabricated using electron beam lithography and standard etching processes, as reported in \citep{Yu2014}, with a single suspended device layer of silicon nitride (SiN). They provide sufficient optical access to enable application of laser cooling and trapping techniques, such as magneto optical trap (MOT) and optical dipole trap in close vicinity of the photonic crystal structures. An overview of our photonic crystal devices is shown in Fig.~\ref{fig:TDCexample}(a), where on-chip waveguides are connected to conventional optical fibers for efficient addressing to the guided modes of the 2D PhC slab by means of a self-collimation scheme \cite{YuPhD,Kosaka1999,Iliewa2004}. This enables direct optical characterization of their properties. The two types of lattice structures to be presented here are: a square lattice of circular holes, as shown in Fig.~\ref{fig:TDCexample}(b), where we explore the anisotropic emission of excited state atoms into the guided photonic modes; and a triangular array of hexagonal holes, depicted in Fig.~\ref{fig:TDCexample}(c), where we will focus on configurations with atomic resonances in the photonic band-gap.

\section*{Engineering atom-photon interactions}

The two photonic crystals slab structures display different regimes for light-matter interactions. For both designed structures, we use numerical tools to investigate how light-matter interactions are affected by the presence of the patterned dielectric. 

We design the photonic crystals by specifying the unit cell geometry, and then computing the band structures using a Finite-Element Method (FEM). The parameter space defining the geometries is explored for useful optical properties such as flat landscape of group velocities and opening of photonic band-gaps, while satisfying practical requirements such as minimum feature sizes and mechanical robustness. The classical electromagnetic Green's tensor is then calculated using Finite-Difference Time-Domain (FDTD) methods on a simulated finite-size photonic crystal slab. Finally, further properties such as the emission characteristics of a dipole near the crystal can be obtained from the Green's tensor \cite{Agarwal1975, Novotny2012, Hung2013}.

\begin{figure*}[htb!]
\centering
\includegraphics[width=1\linewidth]{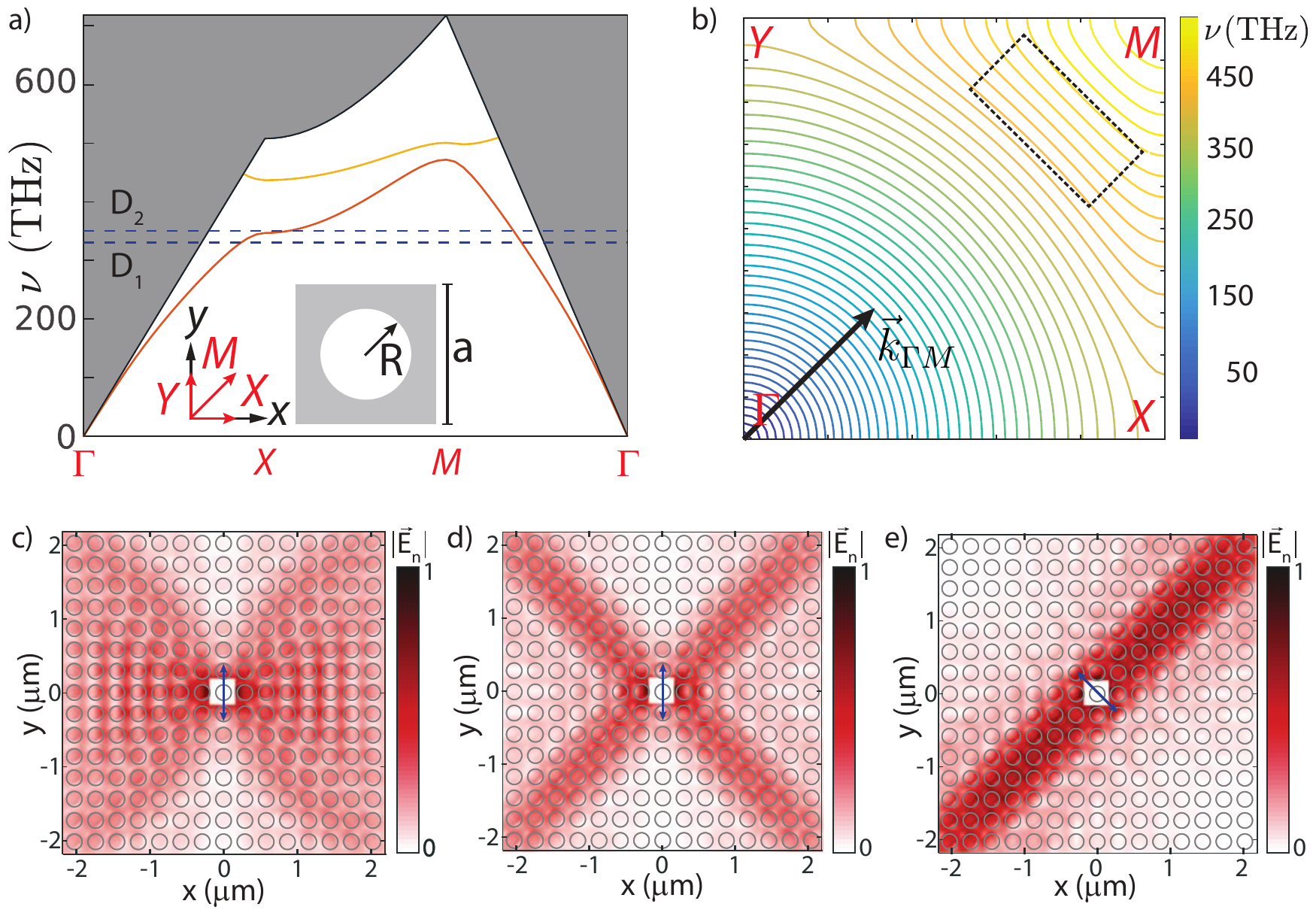}
\caption{Properties of a square lattice of holes in a dielectric slab. (a) The reduced band structures for a unit cell with lattice constant $a=290$~nm, hole radius $R=103$~nm, thickness $t=200$~nm and refractive index $n=2$. Dashed lines mark Cs D$_1$ and D$_2$ transition frequencies. The real space (black arrows) and momentum space (red arrows) basis vectors are shown in the inset. (b) Equi-frequencies curves (EFC) in momentum space for the lowest band shown in (a). The dashed black rectangle shows the region with parallel $\vec{v}_g$, around $\nu = 390\pm20$~THz. The $\vec{k}_{\Gamma M}$ direction is indicated. (c)-(d)-(e) Electric field modulus, $|\vec{\mathrm{E}}_n|$, from a dipole at the center of the unit cell, normalized after removing the field in the immediate vicinity of the dipole, for three different situations. The dipole position and polarization is indicated by the blue arrows. In (c), the dipole frequency is $\nu=320$~THz, the pattern has weak directional features. When the emission frequency is at $\nu=390$~THz, the pattern is clearly directional showing propagation along all $\vec{k}_{\Gamma M}$ directions. In (d), the dipole is polarized along the $y$ direction and emits along both diagonal directions. However, if the dipole is polarized along the diagonal direction, as in (e), only one branch remains.}
\label{fig:sqlatt}
\end{figure*}

The LDOS can be written in terms of the imaginary part of the electromagnetic Green's tensor evaluated at the location of the dipole source itself, $\mathrm{Im}(\textbf{G}(\vec{r},\vec{r},\nu))$. The atomic decay rate from an optically excited state $\ket{j}$ can be expressed in terms of the imaginary Green's tensor as \citep{Hung2013,Agarwal1975,Steck2010,Hughes2015}
\begin{equation}
\Gamma_{\mathrm{Total}}=\frac{8\pi^2\mu_0}{\hbar}\cdot\sum_i \nu_{ij}^2 \mathrm{Tr}\left[ D_{ij} \cdot \mathrm{Im}(\textbf{G}(\vec{r},\vec{r},\nu_{ij})) \right],
\label{equ1}
\end{equation}
where $\nu_{ij}$ stands for the transition frequency between the excited state $\ket{j}$ and the ground state $\ket{g_i}$, $D_{ij}=\bra{g_i}\hat{\vec{d}}^\dagger\ket{j}\bra{j}\hat{\vec{d}}\ket{g_i}$ stands for the transition dipole matrix between the states in consideration, and the summation is over all ground states $\ket{g_i}$. In order to mediate atom-atom interactions using guided mode light, atoms in the vicinity of the photonic structure need to preferentially emit photons into the guided modes of the photonic crystal, instead of emitting into free-space or other loss channels. Such performance can be characterized by the ratio $\Gamma_{\mathrm{2D}}/\Gamma^\prime$, where $\Gamma_{\mathrm{2D}}, \Gamma^\prime$ are the atom decay rate into the guided modes of interest, and into any other channels, respectively, such that $\Gamma_{\mathrm{Total}} = \Gamma_{\mathrm{2D}}+\Gamma^\prime$. It is then useful to design structures that maximize the ratio $\Gamma_{\mathrm{2D}}/\Gamma^\prime$. 

The richness of a 2D structure manifests when studying the spatial profile of the emitted electric field by an atomic dipole near the PhC slab. For example, in the presence of a band-gap, the field produced by the dipole excitation is highly localized but yet inherits the symmetry of the dielectric pattern. If the dipole frequency lies outside a band-gap and thus radiates into the propagating guided modes, highly anisotropic emission can be observed over select frequency ranges \cite{Tudela_PRL2017,Tudela_PRA2017,Tudela2018}. Furthermore, the vector character of the guided mode electric field and the tensor components of the atomic electric dipole operator can further affect the spatial emission pattern.

Our structures were designed to engineer interactions between Cs atoms and the transverse-electric (TE)-like guided modes, where the electric field is polarized predominately along the plane of the slabs. Due to the spin-orbit coupling in the first excited state, Cs has two families of optical transitions, marked by D$_1$ and D$_2$ lines respectively, that can be utilized for optical trapping and studying light-matter interactions. The crystal dimensions are chosen such that the frequencies at high symmetry points are aligned to the Cs D$_1$ and/or D$_2$ transitions at 335~THz (894~nm) and 351~THz (852~nm), respectively. We have thus far constrained our design to be based on a 200~nm thick silicon nitride slab, given its low optical loss at near-infrared range and suitability for lithography and mechanical stability.   

\begin{figure*}[htb!]
\centering
\includegraphics[width=1\linewidth]{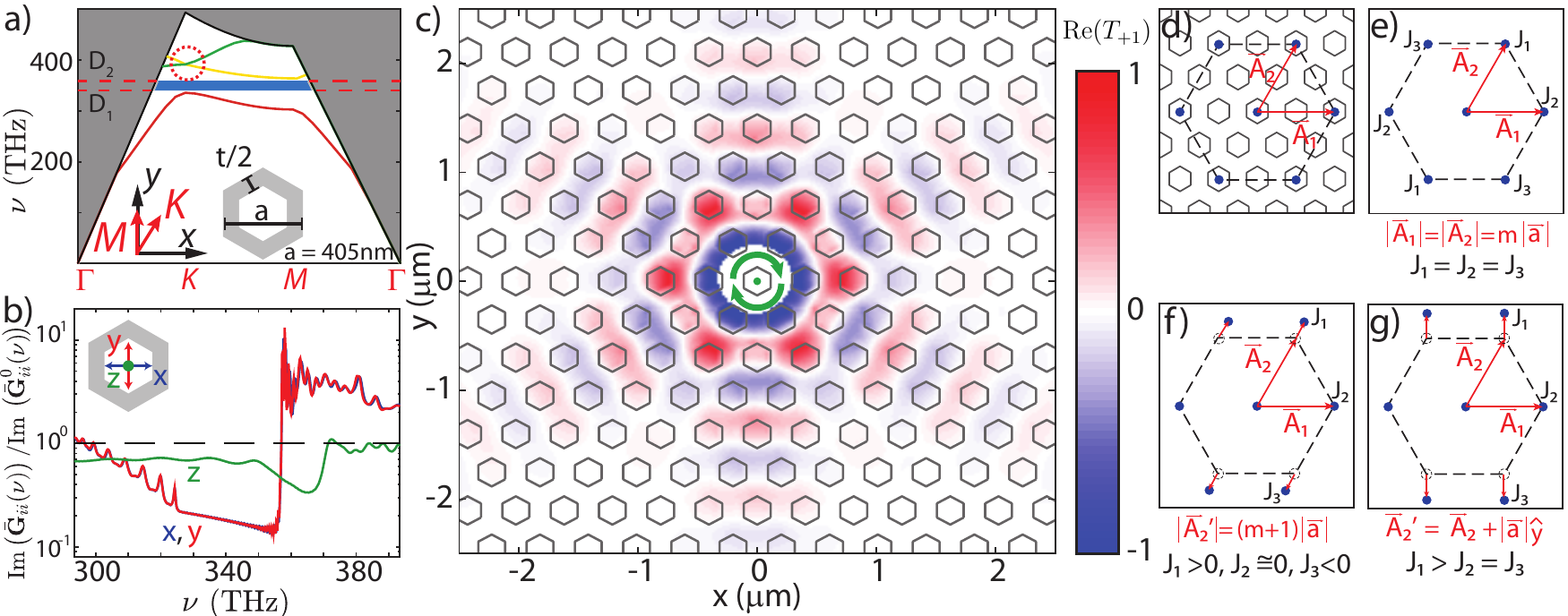}
\caption{Properties of a triangular lattice of hexagonal holes in a dielectric slab (a) The reduced band structure of the hexagonal lattice photonic crystal. A 2D TE-like band-gap (shaded horizontal region) manifests between the top of the lower band (red) and the bottom of the higher bands (green, yellow). The dashed red circle indicates a crossing at the $K$-point between higher bands (see Fig.~\ref{fig:Kpoint}(a)) (b) The imaginary component of the Green's tensor, $\mathrm{Im}(\textbf{G}_{ii}(\vec{r},\vec{r},\nu))/\mathrm{Im}(\textbf{G}^0_{ii}(\vec{r},\vec{r},\nu)$, normalized by the free space tensor components, $i$~= x(blue), y(red), z(green). The excitation dipole position and polarization are shown in the inset. (c) The real part of the Green's tensor component  $\mathrm{Re(T_{+1}}(\vec{r},\nu_0))$, normalized by setting $|\mathrm{Re(T_{+1}}(\vec{r},\nu_0))|$ at the nearest-neighbor cell from the dipole position ($\vec{r}_0$) to 1. The excitation frequency, $\nu_0$, is placed in the middle of the band-gap. The green dot indicates the position of the emitting dipole, while the green arrows indicate its polarization. (d) A super-lattice of atoms associated with lattice vectors $\vec{A}_1, \vec{A}_2$ is formed by placing atoms in selected sites in the photonic crystal. The following interaction parameters $J_i$ can be engineered: (e) $J_1=J_2=J_3$, with $J_i=-0.22<0$ for $m_{11}=m_{22}=m=3$, and $J_i=0.06>0$ for m=4; (f) lengthening the $\vec{A_2}$ vector by one unit lattice vector from the m=4 case results in $J_1=0.06, J_2=-0.01, J_3=-0.06$, which has a weak coupling to the $J_2$ site, forming an effective square-lattice-like interaction; and (g) stretching the super-lattice perpendicularly to $\vec{A_1}$ from the m=2 case creates an anisotropic interaction of $J_1=0.72$ while $J_2=J_3=0.14.$}
\label{fig:HexGrnFnt}
\end{figure*}

\subsection{Anisotropic spontaneous emission in a square lattice photonic crystal slab}

We first consider a PhC slab consisting of a square lattice of circular holes, as shown in Fig.~\ref{fig:TDCexample}(b). The geometry has lattice constant $a=290$~nm, hole radius $R=103$~nm, thickness $t=200$~nm and refractive index $n=2$. This set of parameters was chosen to allow both Cs D$_1$ and D$_2$ lines to couple to a TE-like guided band. Specifically, Cs D$_2$ resonance crosses a region of flat dispersion near the X-point, as shown in Fig.~\ref{fig:sqlatt}(a). In $\vec{k}$-space, the dispersion relation $\nu(\vec{k})$ shows the effects of the dielectric patterning on the PhC as seen in Fig.~\ref{fig:sqlatt}(b), which manifest as equi-frequency curves (EFC) of constant guided mode frequencies. The group velocity $\vec{v}_g = 2\pi \nabla_{\vec{k}} \nu(\vec{k})$ is perpendicular to the EFC that passes through a given $\vec{k}$. As marked in Fig.~\ref{fig:sqlatt}(b), there is a region in $\vec{k}$-space where the EFCs are approximately linear, around $\nu = 390$~THz, indicating that the group velocity points in the same direction (i.e., along $\vec{k}_{\Gamma M}$ as in Fig.~\ref{fig:sqlatt}(b)). Therefore, all excitations with those wave-vectors propagate approximately in the same direction. This gives rise to a self-collimation effect \cite{Witzens2002,MunizPhD} that leads to directional emission \cite{Mekis1999,Tudela_PRL2017,Tudela_PRA2017,Tudela2018,Galve2017}.

In Fig.~\ref{fig:sqlatt}(c-e) we study the effect of engineered band structure on the dipole emission pattern, using the EFC as a guidance. We consider a dipole placed at the center of a hole, polarized parallel to the slab. For a dipole radiating at $\nu=320$~THz and polarized along the $y$ direction, as in Fig.~\ref{fig:sqlatt}(c), the radiation pattern is roughly isotropic and no preferred direction is found. However, if the radiation frequency is at $\nu=390$~THz, as in Fig.~\ref{fig:sqlatt}(d), a clear directional component along $\vec{k}_{\Gamma M}$ is present. Both branches are present due to the dipole polarization and folding symmetry of the square lattice. Finally, we can select a single branch by polarizing the dipole along the diagonal direction, as shown in Fig.~\ref{fig:sqlatt}(e). 

\subsection{Triangular lattice photonic crystal slab}

A photonic crystal with a triangular lattice of hexagonal holes was designed to create a band-gap for the TE-like modes. As indicated in Fig. \ref{fig:HexGrnFnt}(a), we parametrize the photonic crystal unit cell by its lattice constant \textit{a}, and the width \textit{t} of the dielectric tether separating adjacent holes. The unit cell geometry and the band structure for \textit{a}=405~nm and \textit{t}=180~nm are plotted in Fig.~\ref{fig:HexGrnFnt}(a). The TE band-gap spans a frequency range from the K-point at the lower band to the M-point at the higher band, and covers the range of Cs D$_1$ and D$_2$ frequencies. We now study the emission properties of dipole excitation in the vicinity of the photonic crystal slab, by calculating the Green's tensor in the hole center of a unit cell of a photonic crystal slab; see Fig.~\ref{fig:HexGrnFnt}(b). 
The TE band-gap efficiently suppresses emission of a dipole emitter in its frequency range. Numerical simulations show that for the two in-plane polarizations, a suppression of up to 8dB on the spontaneous emission rate can be achieved. There is no significant suppression of decay for the polarization perpendicular to the device plane in the frequency range being considered. This behavior is not reproduced in structures such as the square lattice of holes discussed before, where there is not a complete TE-like band-gap.

The excitation field pattern in the TE-like band-gap is plotted in Fig.~\ref{fig:HexGrnFnt}(c). The modes created by the dipole emitter are non-propagating modes that are confined in the vicinity of the dipole in an evanescent manner. It was purposed that evanescent modes surrounding an atom can be exploited to create a coherent interaction between adjacent atoms with the interaction terms \cite{John1990,Tudela2015,Kurizki1990,Hood2016,Douglas2015}:

\begin{equation}
J_{pq}=\frac{4\pi^2\mu_0\nu_{ij}^2}{\hbar}\cdot \mathrm{Tr}\left[D_{ij} \cdot \mathrm{Re}(\textbf{G}(\vec{r}_p,\vec{r}_q,\nu_{ij}))\right],
\label{equ2}
\end{equation}
where $D_{ij}$ is the dipole matrix as defined previously, and now the Green's Tensor evaluated at the p-th atom position generated by the dipole of the q-th atoms $\textbf{G}(\vec{r}_p,\vec{r}_q)$ is considered. The evanescent guided modes of the photonic crystal mediates atom-atom interactions with the Hamiltonian:

\begin{equation}
\hat{H}_{I}=\sum_{pq}J_{pq}\sigma_{ij}^p\sigma_{ji}^q,
\label{equ3}
\end{equation}
where $\sigma_{ij}^p = \ket{g_i}^p\bra{j}^p$. 

This interaction Hamiltonian forms a coupled system for atoms p and q with transition between the states $\ket{g_i}$ and $\ket{j}$. For simplicity, the following discussion will focus on a two-state atom system. As an example, such two-state system can be realized in our photonic crystal platform with Cs atoms constrained in the $\ket{F=4,m_f=4}$ to $\ket{F'=5, m_f'=5}$ transition. Writing the Green's tensor in its spherical components in Eq(\ref{equ2}), only the term $T_{+1}(\vec{r},\nu)=(\hat{e}_+\cdot\textbf{G}\left(\vec{r},\vec{r}_0,\nu\right)\cdot\hat{e}_-)$ is non-zero through the trace operation, where $\hat{e}_\pm=\frac{\hat{x}\pm i \hat{y}}{\sqrt{2}}$ and $\vec{r}_0$ is the position of the emitting dipole. The term is calculated numerically from the counter-clockwise circular component of electric field $\hat{e}_+\cdot\vec{E}$ generated by a clockwise circular dipole source $p\hat{e}_-$ placed at the origin, $\vec{r}_0$. The profile of $\mathrm{Re(T_{+1}}(\vec{r},\nu_0))$, with $\nu_0$ within the bandwidth of the 2D TE band-gap, is plotted in Fig.\ref{fig:HexGrnFnt}(c).

The geometric pattern of the Green's tensor $\mathrm{Re}(\textbf{G})$ can be exploited to engineer the form of the interaction Hamiltonian. The relative position of atoms on the photonic crystal determines the sign and strength of the atom-atom interaction as facilitated by the evanescent field pattern. Assuming a site-wise controlled placement of atoms can be realized with technique like the optical tweezers \citep{Eriksen2002}, atoms can be assembled into a super-lattice on the photonic crystal to form a quantum many-body system with desired atom-atom interaction. We define such a super-lattice by lattice vectors spanned by integer multiple combinations of the base lattice vectors, $\vec{A_i}=m_{ij}\cdot\vec{a_j}$. In Fig.~\ref{fig:HexGrnFnt}(d)-(g) we investigate this capability by identifying configurations that would create nearest-neighbor coupling of controlled sign and engineered anisotropy. 

\begin{figure}[htb!]
\centering
\includegraphics[width=1\linewidth]{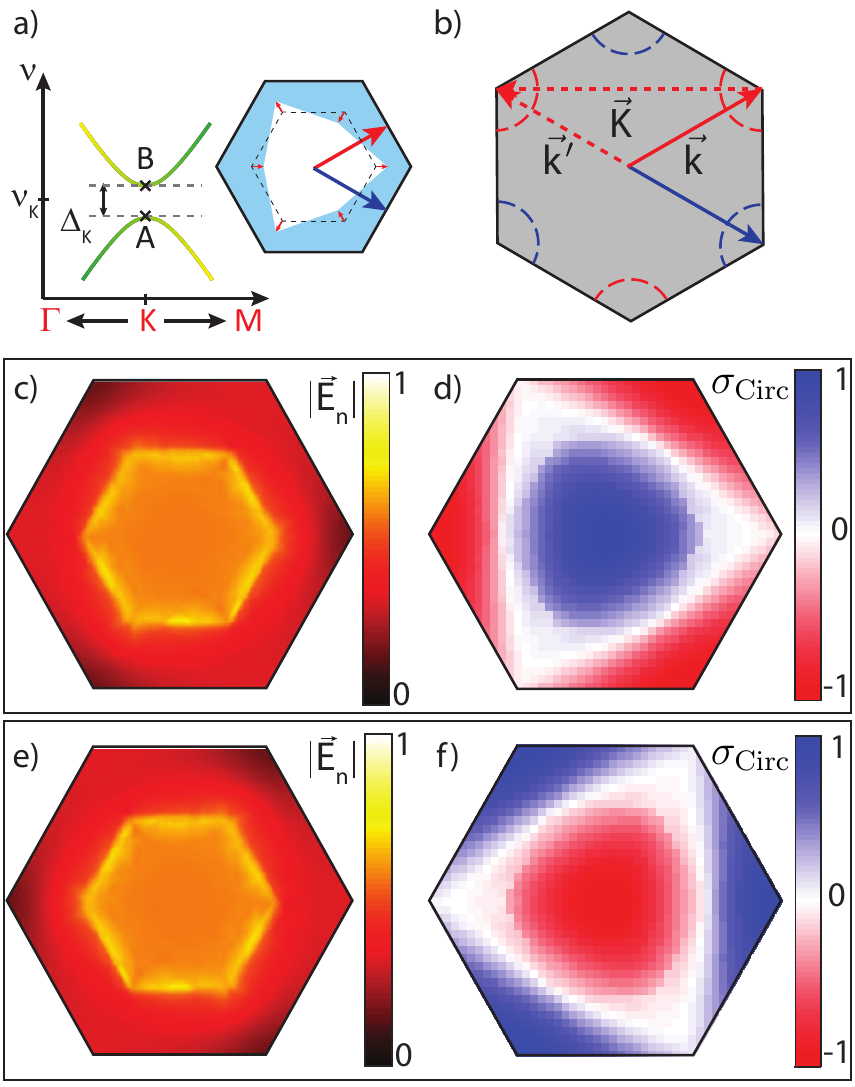}
\caption{Lifting the degeneracy at the $K$-point. (a) A modified unit cell for the hexagonal photonic crystal where the vertices of the hole are moved to break the mirror-symmetry in the $K$-direction while maintaining 3-fold rotation symmetry around its center. Such deformation lifts the degeneracy of the two upper bands at the $K$ point plotted in Fig.~\ref{fig:HexGrnFnt}(a). (b) A guided mode of the photonic crystal with $\vec{k}$ lying on the $K$-point of the Brillouin zone maintains the 3-fold rotation symmetry, as such rotation operation transforms the momentum vector onto $\vec{k'}=\vec{k}+\vec{K}$, where $\vec{K}$ is a lattice momentum, and $\vec{k}$ and $\vec{k'}$ are equivalent. (c) and (d) show the electric field pattern $|\vec{E}|$ and circular polarization ratio, $\sigma_{Circ}$ for point A in the dispersion plot in (a), for a unit cell with vertices shifted by 2.5~nm, lifting the degeneracy at $\nu_{K}=387.6$~THz by $\Delta_K=0.7$~THz. (e) and (f) show the corresponding plots for point B. We note that the modes have strong electric field in the center of the hole with purely circular polarization.}
\label{fig:Kpoint}
\end{figure}

It is worth noting that the band structure of the triangular PhC enables reliable creation of symmetry-protected, purely circularly polarized mode pockets in a hexagonal lattice structure. The mode crossing in the two upper bands at the $K$ point, encircled in red on Fig.~\ref{fig:HexGrnFnt}(a), is a degeneracy that is protected by the six-fold rotation symmetry of the lattice \cite{Wen2008} around the Z axis perpendicular to the device plane. This degeneracy can be lifted by perturbing the unit cell dielectric distribution, while preserving a 3-fold rotation symmetry, as shown in Fig.~\ref{fig:Kpoint}(a)-(b). At high-symmetry points, such as the center of the hole, the guided mode field pattern is necessarily purely circularly polarized. Any defined in-plane linear direction is incompatible with the 3-fold rotation symmetry. In fig. ~\ref{fig:Kpoint}(c)-(f) we plot the field profiles and their circular polarization fraction, defined as $\sigma_{Circ}=\frac{I_\mathrm{CCW}-I_\mathrm{CW}}{I_\mathrm{CCW}+I_\mathrm{CW}}$, where $I_\mathrm{CW}, I_\mathrm{CCW}$ are the local intensity of clockwise and counter-clockwise polarizations. To respect time-reversal symmetry, the circularly polarized guided modes are necessarily chiral \cite{Lodahl2017}, with the locking of spin (defined by the polarization) and momentum (defined by the propagation direction). By selectively coupling the guided modes with trapped atoms polarized at certain magnetic sub-level $m_f$, one can potentially create topological properties while using a linear dielectric media \citep{Perczel_PRL2017,Shi2017}.

\begin{figure}[htb!]
\centering
\includegraphics[width=1\linewidth]{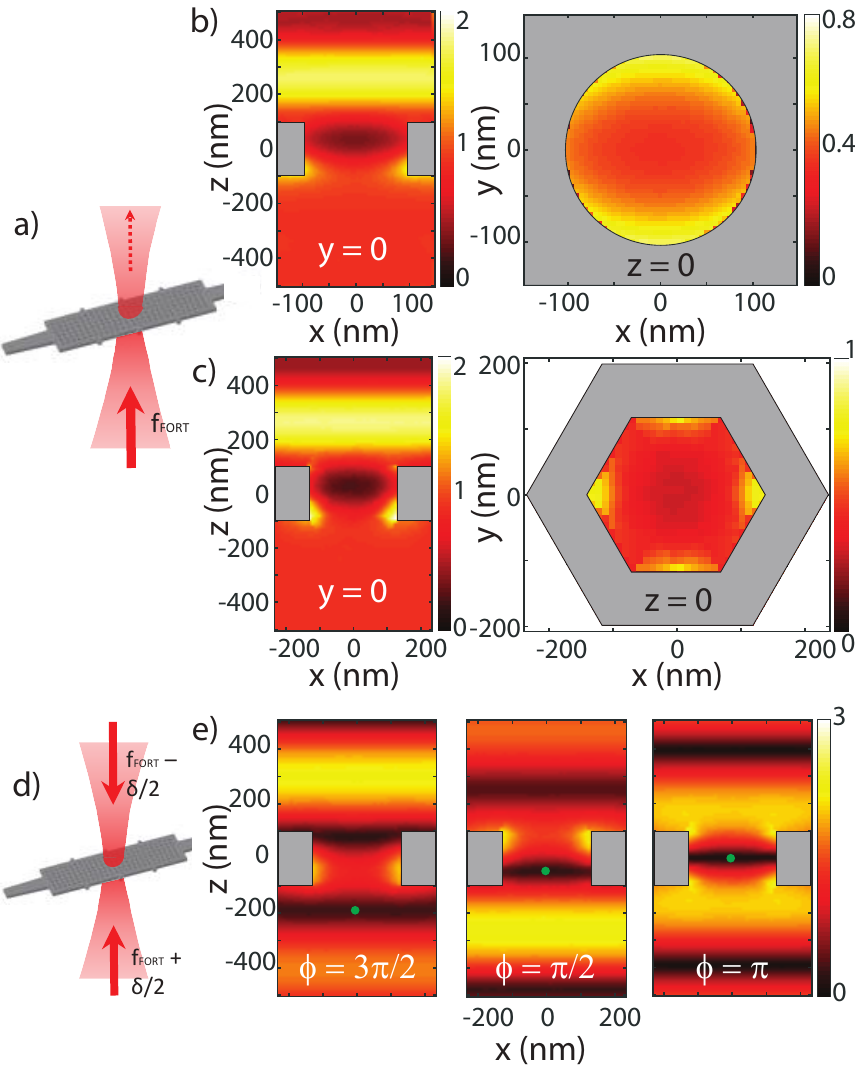}
\caption{Optical dipole traps near the PhC slab. (a) A linearly polarized beam is sent with wavevector k normal to the plane of the slab, forming local intensity minima for both the square lattice (b) and the hexagonal lattice (c) structures. Left and right panels show intensity cross-sections in the $x-z$ and $x-y$ planes, for $y = 0$ and $ z = 0 $, respectively, as indicated. (d) An optical conveyor belt can be formed by using near detuned counter-propagating beams.(e) For different relative phases between the beams the intensity pattern displaces its intensity minimum, green circles, into the vacuum spaces. Here the triangular structure is depicted. Relative phases indicated in each panel.}
\label{fig:TDCtrap}
\end{figure}

\section*{Optical Trapping Schemes}
Here we present an optical loading and trapping scheme for atoms in close proximity of the 2D photonic crystal. The method is based on a blue-detuned optical `conveyor belt' technique \citep{Schrader2001,Tudela2015,MunizPhD,Burgers2018}; see Fig.~\ref{fig:TDCtrap}. The calculations have been carried out for both structures described earlier. First, we consider the case of a single linearly polarized beam, near-detuned with respect to the Cs D$_2$ wavelength, incident normally to the slab plane, as in Fig.~\ref{fig:TDCtrap}(a). The reflection from the dielectric creates a strong intensity modulation as shown in Fig.~\ref{fig:TDCtrap}(b)-(c). Near the center of the hole, a local intensity minimum is present and can be utilized to localize cold atoms. It is also possible to trap above the structure on several intensity maxima using red detuned light.

The blue-detuned trap is favored over conventional red-detuned trap due to the fact that optical intensity tends to sharply increase near sub-wavelength features in the dielectric structure as a result of near-field effects. A blue-detuned trap with such intensity pattern would then create a strong potential barrier near the dielectric surface to prevent un-trapped atoms from crashing into the dielectric structures \cite{Tudela2015,MunizPhD}. The same mechanism also prevents trapped atoms from coming into close vicinity of dielectric surfaces, where the Casimir-Polder force \citep{Hung2013,Tudela2015,Buhmann2007} becomes dominant. Cs accumulation on the devices was found to be the primary limitation of device lifetime in previous experiments with cesium near nano-structures \citep{Goban2014,Goban2015}. We suggest that this could be alleviated with blue-detuned traps.

To perform an optical conveyor-belt, a moving 1D optical lattice can be formed using two linearly polarized, counter-propagating laser beams, with a relative detuning $\delta$ and incident normally with respect to the device plane, as shown in Fig.~\ref{fig:TDCtrap}(d). The lattice can be loaded with cold atoms from a conventional free-space MOT sufficiently far from the nano-structure, and can be continuously moved into and through the device layer at a rate controlled by the detuning $\delta$. To provide transverse confinement in such a blue detuned 1D lattice, additional beams can be used that do not alter this discussion significantly. Figure ~\ref{fig:TDCtrap}(e) shows the trap pattern for different relative phases between the beams. Confinement in the additional $y$ direction is guaranteed as seen in $xy$ patterns in Fig.~\ref{fig:TDCtrap}(b)-(c). It is possible to extract information of the phase of the moving lattice, hence inferring the position of the loaded atoms, by imaging the scattering of the device layer or collecting light from the device through an out-coupling port \citep{Burgers2018}. Moreover, since a single incident beam can already form a local trap intensity minimum within the hole, it is possible to abruptly turn off one of the two beams as the intensity node of the blue lattice passes through the device layer to convert the conveyor-belt directly into a localized trap. Further schemes using guided modes to create traps can be investigated, as suggested in \cite{Tudela2015}.

Using optical tweezers in 2D PhC systems is a powerful alternative. Recent development of optical tweezers techniques as used in \citep{Lester2015,Thompson2013, Endres2016,Barredo2016, Kim2018} allows for precise and dynamic placement of multiple atoms. The optical access of our chip configuration, as well as the planar geometry of the 2D photonic crystal slab, are compatible with optical tweezers techniques. In the near term, controlled placement of pairs of atoms in determined relative positions could be achieved, as a mean to directly characterize the anisotropic interactions described previously. Ultimately, an array of atoms could be arranged in a super-lattice on the photonic crystal with multiple optical tweezers to engineer quantum many-body systems with controlled interactions.

\begin{figure}[htb!]
\centering
\includegraphics[width=1\linewidth]{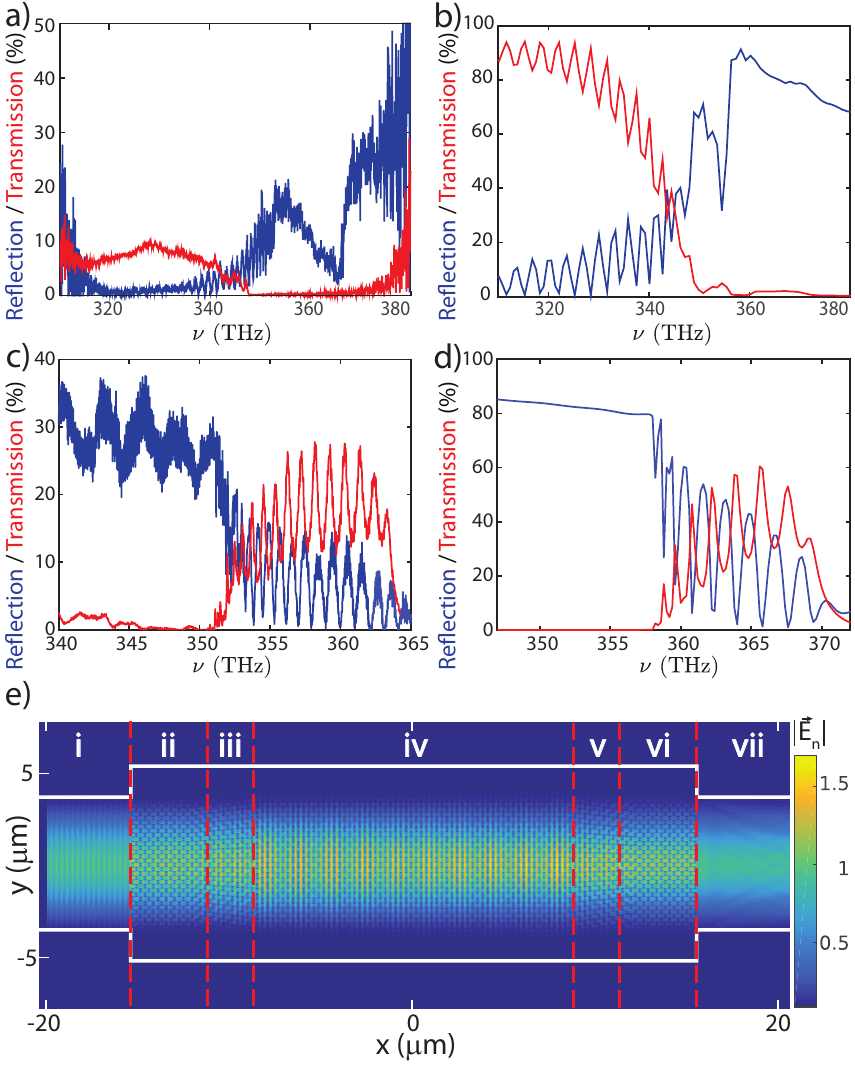}
\caption{Device characterization. For a square lattice, the comparison between measured (a) and simulated (b) reflection and transmission traces revels some common features. For the hexagonal lattice, the measured (c) and simulated (d) traces reflects the existence of a band-gap. We note that the total device lengths in the simulated devices are shorter, due to computational resource limits, resulting in a larger frequency spacing between the resonances. The small shift in resonance frequencies between reflection and transmission traces in (c) resulted from an oxygen plasma cleaning process carried out between the two measurements. (e) Simulated field intensity for a incident guided mode, $|\vec{\mathrm{E}}_n|$, coming through the left waveguide and propagating through the crystal near the Cs D$_2$ line, white lines indicate the device contour. Multiple photonic sections are divided by the dashed lines. Section i and vii are the input/output 8$\mu m$ rectangular waveguides; sections ii and vi are the self collimation regions with the $\Gamma$-$M$ direction aligned along the horizontal axis; sections iii and v are the transition sections to align the self-collimated light at Cs D2 frequency to the M-point of the PhC in section iv. Note the incoming waveguide mode does not diverge as it enters the crystal from the left.}
\label{fig:SimVsMeas}
\end{figure}

\section*{Device characterization}

We fabricate the 2D PhC devices using a similar process to that presented in Ref.~\citep{Yu2014}. The photonic crystal slab is suspended atop a through-window chemically etched through the silicon substrate, and is connected to the chip substrate via a set of silicon nitride waveguides and tethers. The waveguides can be efficiently coupled to conventional single-mode optical fibers using the method developed in Ref.~\citep{Cohen2013}. We then connect these waveguides of 500~nm width to the 2D photonic crystal slab by using a 1:10 linear taper to gradually enlarge the waveguide width to 8~$\mu$m. The function of the linear taper is two-fold: Optically, it widens the optical mode profile from sub-wavelength to several $\mu$m. As the guided mode enters the 2D structure, the widened mode profile suppresses the spread in transverse k component of the optical mode, so that it can couple into a 2D photonic crystal mode of well-defined lattice momentum vector. Mechanically, the widening taper distributes the high tensile stress carried on the waveguide evenly onto the larger, more rigid 2D photonic crystal slab, hence improving device yield. We have fabricated fully-suspended 2D PhC slabs with a size up to 30~$\mu$m by 90~$\mu$m with high ($>$90\%) yield \citep{YuPhD}.

For the 2D system in consideration, optical input can be introduced into the crystal from a continuum of directions in the device plane. This is achieved in our system by maintaining the input waveguide configuration, but rotating the lattice vectors of the 2D PhC. Typically, each chip contains 16 devices, and a gradual rotation of the crystal lattices across the devices allows studying the device's response to input light in quasi-continuous directions. We found it to be most informative to characterize these devices from the high-symmetry directions of the Brillouin zone, as this provides direct information of the frequencies of the upper and lower bands at these directions. For the case of the hexagonal crystal, the TE-like band-gap exists between the frequencies of the lower band of the $K$-point and the higher band of the $M$-point. Direct measurement of these frequencies allows us to infer the frequency range of the 2D photonic band-gap.

A set of measured device transmission and reflection spectra is plotted on Fig.~\ref{fig:SimVsMeas}, with FDTD simulation traces for comparison. The simulations of devices were carried out using identical photonic crystal parameters, but with reduced device size, due to limits of computational resources. Spectral features such as stop-bands and resonance dips can be identified in simulation and measured traces with good correspondence. We note that the rapid interference fringes on the measured spectra result from residual reflections from connecting waveguides and fiber coupling structures. These fringes are not intrinsic to the photonic crystals, and can be alleviated using a single row of holes that provide an effective AR-coating \citep{Lee08,MunizPhD}. In addition to observing the band-edges of the high-symmetry directions, useful information regarding the PhC can be inferred from the various features on the transmission and reflection spectra.

We will focus our attention on the band-edges, where the frequency of input light changes from propagation band to band-gap. Figure \ref{fig:SimVsMeas}(a)-(b) show the comparison between measurements and numerical simulations, respectively, for the lower band-edge at the $X$-point of the square lattice. The TE band-gap in the $X$ orientation prevents transmission of the TE-like mode above about 348~THz. Also, the reflection increases rapidly above this frequency. 

Panels (c)-(d) in Fig.~\ref{fig:SimVsMeas} depict transmission and reflection spectra of a hexagonal photonic crystal in the $M$ direction. The TE band-gap of the hexagonal photonic crystal creates an abrupt suppression of transmission. We observe a set of resonances with reducing frequency spacing as the frequency approaches the band-edge. These resonances result from reflection at the boundaries of the photonic crystal, and the reduction of resonance spacing resulted from increasing group index of the guided mode, similar to the 1D case \citep{Yu2014}. The reduction of contrast of the resonances as the frequency approaches the band-edge results also from the increased group index, which creates enhanced field build-up inside the photonic crystal and enhances scattering into other modes of the 2D photonic crystal slab \cite{Tanaka2004}.

Furthermore, from these numerical simulations it is possible to study the field profile as it propagates along the device as in Fig.~\ref{fig:SimVsMeas}(e). In order to avoid a divergence of the input mode, the nominal PhC is placed in between two self-collimation sections where the unit cells are aligned such that at the frequencies of interest, the group velocity points along the left-right direction as discussed before \citep{Witzens2002,MunizPhD}. The nominal section in this case consists of a unit cell lattice constant of 360nm and a hole radius of 105nm. These dimensions place the Cs D$_2$ frequency near the $M$-point of the lower band.

\section*{Outlook}
In this article, we have described a collection of components and capabilities that will allow experiments with cold atoms in 2D photonic crystal system \cite{Tudela2015,Hung2016}. Photonic crystals can indeed be built to produce desired band structures near cesium resonances. Corresponding devices have been fabricated reliably using well-understood processes, and the expected optical features in reflection and transmission spectra have been verified \cite{YuPhD}. The Green's tensor calculations show that a wide variety of atom-light interaction can be engineered, including enhancement and suppression of decay rates, directional propagation, and finite-range atom-atom interactions mediated by photons. An optical trapping and loading scheme has been proposed to allow placement of atoms in positions in the crystal structures to demonstrate the suggested phenomena. In the foreseeable future, we should take several steps to further verify the properties of the devices with atoms. The loading and trapping of atoms using the optical conveyor belt should allow demonstration of enhanced or suppressed atomic decay rates corresponding to the Green's tensor results, which should be measurable with time-domain decay rate measurements with the integrated waveguides on the chip \cite{Goban2015}. Optical tweezers should then be deployed in our system to directly measure the predicted anisotropic interactions. Achieving these steps sequentially should steadily bring us toward realization of engineer-able quantum many-body system in our 2D photonic crystal devices.

\section*{Material and Methods}
The photonic crystal band structures are simulated using \textit{COMSOL Multiphysics 3.5a} finite-element eigenvalue solver \cite{Comsol}. A unit cell geometry is plotted, a Bloch boundary condition with lattice momentum $\vec{k}$ is specified for each simulation, and the eigen-frequencies are solved for each $\vec{k}$ along the high-symmetry directions to form the band structures.

The Green's tensor calculations are carried out using the \textit{MEEP} package \cite{Oskoi2010}, and also \textit{Lumerical FDTD} software \cite{Lumerical}. A finite-size photonic crystal slab is created in a 3D simulation volume, and a dipole source is placed in a desired location to provide excitation. For the decay rate enhancement, the imaginary component of the resulting field is recorded at the dipole position, and Fourier transform is carried out to extract the frequency response. The resulting emission mode profile from the dipole source is also recorded for the anisotropic emission studies.

The optical trapping conveyor scheme is designed also with \textit{COMSOL Multiphysics 3.5a}, where a single-frequency plane wave is injected onto a unit cell from perpendicular orientation to compute the resulting field pattern. The electric field is then exported, parity-reversed to create counter-propagating wave, and superimposed with the forward-propagating mode to form a standing wave. A relative phase between the two modes is swept to create the conveyor motion.

The fabrication of devices starts with 200$\mu$m thickness silicon substrates pre-coated with 200nm stoichiometric silicon nitride on both sides by \textit{Silicon Valley Microelectronics}. The waveguides and photonic crystals were defined using a \textit{Raith EBPG 5000+} lithography tool with \textit{ZEON Chemicals ZEP520A} electron beam resist, and pattern-transferred into silicon nitride device layer using reactive ion etching with a \textit{Oxford PlasmaLab 100} tool with $C_4F_8$ and $SF_6$ chemistry. The silicon substrate through-hole etch is done using potassium hydroxide solution, and subsequent cleaning with \textit{CyanTek Nanostrip} and 1:10 buffered hydrofluoric acid. The finished devices are brought out of liquid using critical point drying. The optical testing of devices is done by injecting broad-band light from a \textit{InPhenix SLED} light source into the device with cleaved 780HP optical fiber, and the reflection spectrum is analyzed using an \textit{Anritsu MS9740A} optical spectrum analyzer.


\begin{acknowledgments}
We gratefully acknowledge discussions with Andrew McClung, Jonathan Hood, Lucas Peng, Alexander Burgers, Michael Martin and Ana Asenjo-Garc\'ia from the Caltech Quantum Optics Group, and with Alejandro Gonz\'alez-Tudela and Ignacio Cirac (MPQ, Garching). S.-P. Y. and J. A. M. acknowledge support from the International Fulbright Science and Technology Award. C.-L. H. acknowledges support from the Air Force Office of Scientific Research (AFOSR), Grant No. FA9550-17-1-0298 and the Office of Naval Research (ONR), Grant No. N00014-17-1-2289. HJK acknowledges funding from ONR Grant No. N00014-16-1-2399, ONR MURI Quantum Opto-Mechanics with Atoms and Nanostructured Diamond Grant No. N00014-15-1-2761, AFOSR MURI Photonic Quantum Matter Grant No. FA9550-16-1-0323, the National Science Foundation (NSF) Grant No. PHY-1205729, and the NSF Institute for Quantum Information and Matter Grant No. PHY-1125565, as well as the support of the Caltech Kavli Nanoscience Institute (KNI) and the cleanroom facilities of O. Painter and his group, where device fabrication was carried out.
\end{acknowledgments}

\bibliography{TDC_ref}

\begin{thebibliography}{67}%
\makeatletter
\providecommand \@ifxundefined [1]{%
 \@ifx{#1\undefined}
}%
\providecommand \@ifnum [1]{%
 \ifnum #1\expandafter \@firstoftwo
 \else \expandafter \@secondoftwo
 \fi
}%
\providecommand \@ifx [1]{%
 \ifx #1\expandafter \@firstoftwo
 \else \expandafter \@secondoftwo
 \fi
}%
\providecommand \natexlab [1]{#1}%
\providecommand \enquote  [1]{``#1''}%
\providecommand \bibnamefont  [1]{#1}%
\providecommand \bibfnamefont [1]{#1}%
\providecommand \citenamefont [1]{#1}%
\providecommand \href@noop [0]{\@secondoftwo}%
\providecommand \href [0]{\begingroup \@sanitize@url \@href}%
\providecommand \@href[1]{\@@startlink{#1}\@@href}%
\providecommand \@@href[1]{\endgroup#1\@@endlink}%
\providecommand \@sanitize@url [0]{\catcode `\\12\catcode `\$12\catcode
  `\&12\catcode `\#12\catcode `\^12\catcode `\_12\catcode `\%12\relax}%
\providecommand \@@startlink[1]{}%
\providecommand \@@endlink[0]{}%
\providecommand \url  [0]{\begingroup\@sanitize@url \@url }%
\providecommand \@url [1]{\endgroup\@href {#1}{\urlprefix }}%
\providecommand \urlprefix  [0]{URL }%
\providecommand \Eprint [0]{\href }%
\providecommand \doibase [0]{http://dx.doi.org/}%
\providecommand \selectlanguage [0]{\@gobble}%
\providecommand \bibinfo  [0]{\@secondoftwo}%
\providecommand \bibfield  [0]{\@secondoftwo}%
\providecommand \translation [1]{[#1]}%
\providecommand \BibitemOpen [0]{}%
\providecommand \bibitemStop [0]{}%
\providecommand \bibitemNoStop [0]{.\EOS\space}%
\providecommand \EOS [0]{\spacefactor3000\relax}%
\providecommand \BibitemShut  [1]{\csname bibitem#1\endcsname}%
\let\auto@bib@innerbib\@empty
\bibitem [{\citenamefont {Chang}\ \emph {et~al.}(2018)\citenamefont {Chang},
  \citenamefont {Douglas}, \citenamefont {Gonz\'alez-Tudela}, \citenamefont
  {Hung},\ and\ \citenamefont {Kimble}}]{Chang2018}%
  \BibitemOpen
  \bibfield  {author} {\bibinfo {author} {\bibfnamefont {D.~E.}\ \bibnamefont
  {Chang}}, \bibinfo {author} {\bibfnamefont {J.~S.}\ \bibnamefont {Douglas}},
  \bibinfo {author} {\bibfnamefont {A.}~\bibnamefont {Gonz\'alez-Tudela}},
  \bibinfo {author} {\bibfnamefont {C.-L.}\ \bibnamefont {Hung}}, \ and\
  \bibinfo {author} {\bibfnamefont {H.~J.}\ \bibnamefont {Kimble}},\ }\bibfield
   {title} {\enquote {\bibinfo {title} {Colloquium: Quantum matter built from
  nanoscopic lattices of atoms and photons},}\ }\href {\doibase
  10.1103/RevModPhys.90.031002} {\bibfield  {journal} {\bibinfo  {journal}
  {Rev. Mod. Phys.}\ }\textbf {\bibinfo {volume} {90}},\ \bibinfo {pages}
  {031002} (\bibinfo {year} {2018})}\BibitemShut {NoStop}%
\bibitem [{\citenamefont {Goban}\ \emph {et~al.}(2012)\citenamefont {Goban},
  \citenamefont {Choi}, \citenamefont {Alton}, \citenamefont {Ding},
  \citenamefont {Lacro{\^u}te}, \citenamefont {Pototschnig}, \citenamefont
  {Thiele}, \citenamefont {Stern},\ and\ \citenamefont {Kimble}}]{Goban2012}%
  \BibitemOpen
  \bibfield  {author} {\bibinfo {author} {\bibfnamefont {A.}~\bibnamefont
  {Goban}}, \bibinfo {author} {\bibfnamefont {K.~S.}\ \bibnamefont {Choi}},
  \bibinfo {author} {\bibfnamefont {D.~J.}\ \bibnamefont {Alton}}, \bibinfo
  {author} {\bibfnamefont {D.}~\bibnamefont {Ding}}, \bibinfo {author}
  {\bibfnamefont {C.}~\bibnamefont {Lacro{\^u}te}}, \bibinfo {author}
  {\bibfnamefont {M.}~\bibnamefont {Pototschnig}}, \bibinfo {author}
  {\bibfnamefont {T.}~\bibnamefont {Thiele}}, \bibinfo {author} {\bibfnamefont
  {N.~P.}\ \bibnamefont {Stern}}, \ and\ \bibinfo {author} {\bibfnamefont
  {H.~J.}\ \bibnamefont {Kimble}},\ }\bibfield  {title} {\enquote {\bibinfo
  {title} {Demonstration of a state-insensitive, compensated nanofiber trap},}\
  }\href@noop {} {\bibfield  {journal} {\bibinfo  {journal} {Physical Review
  Letters}\ }\textbf {\bibinfo {volume} {109}},\ \bibinfo {pages} {033603}
  (\bibinfo {year} {2012})}\BibitemShut {NoStop}%
\bibitem [{\citenamefont {Sagu\'e}\ \emph {et~al.}(2007)\citenamefont
  {Sagu\'e}, \citenamefont {Vetsch}, \citenamefont {Alt}, \citenamefont
  {Meschede},\ and\ \citenamefont {Rauschenbeutel}}]{Sague_2007}%
  \BibitemOpen
  \bibfield  {author} {\bibinfo {author} {\bibfnamefont {G.}~\bibnamefont
  {Sagu\'e}}, \bibinfo {author} {\bibfnamefont {E.}~\bibnamefont {Vetsch}},
  \bibinfo {author} {\bibfnamefont {W.}~\bibnamefont {Alt}}, \bibinfo {author}
  {\bibfnamefont {D.}~\bibnamefont {Meschede}}, \ and\ \bibinfo {author}
  {\bibfnamefont {A.}~\bibnamefont {Rauschenbeutel}},\ }\bibfield  {title}
  {\enquote {\bibinfo {title} {Cold-atom physics using ultrathin optical
  fibers: Light-induced dipole forces and surface interactions},}\ }\href
  {\doibase 10.1103/PhysRevLett.99.163602} {\bibfield  {journal} {\bibinfo
  {journal} {Phys. Rev. Lett.}\ }\textbf {\bibinfo {volume} {99}},\ \bibinfo
  {pages} {163602} (\bibinfo {year} {2007})}\BibitemShut {NoStop}%
\bibitem [{\citenamefont {Vetsch}\ \emph {et~al.}(2010)\citenamefont {Vetsch},
  \citenamefont {Reitz}, \citenamefont {Sagu\'e}, \citenamefont {Schmidt},
  \citenamefont {Dawkins},\ and\ \citenamefont {Rauschenbeutel}}]{Vetsch2010}%
  \BibitemOpen
  \bibfield  {author} {\bibinfo {author} {\bibfnamefont {E.}~\bibnamefont
  {Vetsch}}, \bibinfo {author} {\bibfnamefont {D.}~\bibnamefont {Reitz}},
  \bibinfo {author} {\bibfnamefont {G.}~\bibnamefont {Sagu\'e}}, \bibinfo
  {author} {\bibfnamefont {R.}~\bibnamefont {Schmidt}}, \bibinfo {author}
  {\bibfnamefont {S.~T.}\ \bibnamefont {Dawkins}}, \ and\ \bibinfo {author}
  {\bibfnamefont {A.}~\bibnamefont {Rauschenbeutel}},\ }\bibfield  {title}
  {\enquote {\bibinfo {title} {Optical interface created by laser-cooled atoms
  trapped in the evanescent field surrounding an optical nanofiber},}\ }\href
  {\doibase 10.1103/PhysRevLett.104.203603} {\bibfield  {journal} {\bibinfo
  {journal} {Phys. Rev. Lett.}\ }\textbf {\bibinfo {volume} {104}},\ \bibinfo
  {pages} {203603} (\bibinfo {year} {2010})}\BibitemShut {NoStop}%
\bibitem [{\citenamefont {Corzo}\ \emph {et~al.}(2016)\citenamefont {Corzo},
  \citenamefont {Gouraud}, \citenamefont {Chandra}, \citenamefont {Goban},
  \citenamefont {Sheremet}, \citenamefont {Kupriyanov},\ and\ \citenamefont
  {Laurat}}]{Corzo2016}%
  \BibitemOpen
  \bibfield  {author} {\bibinfo {author} {\bibfnamefont {N.~V.}\ \bibnamefont
  {Corzo}}, \bibinfo {author} {\bibfnamefont {B.}~\bibnamefont {Gouraud}},
  \bibinfo {author} {\bibfnamefont {A.}~\bibnamefont {Chandra}}, \bibinfo
  {author} {\bibfnamefont {A.}~\bibnamefont {Goban}}, \bibinfo {author}
  {\bibfnamefont {A.~S.}\ \bibnamefont {Sheremet}}, \bibinfo {author}
  {\bibfnamefont {D.~V.}\ \bibnamefont {Kupriyanov}}, \ and\ \bibinfo {author}
  {\bibfnamefont {J.}~\bibnamefont {Laurat}},\ }\bibfield  {title} {\enquote
  {\bibinfo {title} {Large bragg reflection from one-dimensional chains of
  trapped atoms near a nanoscale waveguide},}\ }\href {\doibase
  10.1103/PhysRevLett.117.133603} {\bibfield  {journal} {\bibinfo  {journal}
  {Physical Review Letters}\ }\textbf {\bibinfo {volume} {117}},\ \bibinfo
  {pages} {133603} (\bibinfo {year} {2016})}\BibitemShut {NoStop}%
\bibitem [{\citenamefont {JD}\ \emph {et~al.}(2013)\citenamefont {JD},
  \citenamefont {TG}, \citenamefont {de~Leon~NP}, \citenamefont {J},
  \citenamefont {AV}, \citenamefont {M}, \citenamefont {AS}, \citenamefont
  {V},\ and\ \citenamefont {MD}}]{Thompson2013}%
  \BibitemOpen
  \bibfield  {author} {\bibinfo {author} {\bibfnamefont {T.}~\bibnamefont
  {JD}}, \bibinfo {author} {\bibfnamefont {T.}~\bibnamefont {TG}}, \bibinfo
  {author} {\bibnamefont {de~Leon~NP}}, \bibinfo {author} {\bibfnamefont
  {F.}~\bibnamefont {J}}, \bibinfo {author} {\bibfnamefont {A.}~\bibnamefont
  {AV}}, \bibinfo {author} {\bibfnamefont {G.}~\bibnamefont {M}}, \bibinfo
  {author} {\bibfnamefont {Z.}~\bibnamefont {AS}}, \bibinfo {author}
  {\bibfnamefont {V.}~\bibnamefont {V}}, \ and\ \bibinfo {author}
  {\bibfnamefont {L.}~\bibnamefont {MD}},\ }\bibfield  {title} {\enquote
  {\bibinfo {title} {Coupling a single trapped atom to a nanoscale optical
  cavity},}\ }\href {\doibase 10.1126/science.1237125} {\bibfield  {journal}
  {\bibinfo  {journal} {Science}\ }\textbf {\bibinfo {volume} {340}},\ \bibinfo
  {pages} {1202--5} (\bibinfo {year} {2013})}\BibitemShut {NoStop}%
\bibitem [{\citenamefont {Alton}\ \emph {et~al.}(2011)\citenamefont {Alton},
  \citenamefont {Stern}, \citenamefont {Aoki}, \citenamefont {Lee},
  \citenamefont {Ostby}, \citenamefont {Vahala},\ and\ \citenamefont
  {Kimble}}]{Alton2010}%
  \BibitemOpen
  \bibfield  {author} {\bibinfo {author} {\bibfnamefont {D.~J.}\ \bibnamefont
  {Alton}}, \bibinfo {author} {\bibfnamefont {N.~P.}\ \bibnamefont {Stern}},
  \bibinfo {author} {\bibfnamefont {T.}~\bibnamefont {Aoki}}, \bibinfo {author}
  {\bibfnamefont {H.}~\bibnamefont {Lee}}, \bibinfo {author} {\bibfnamefont
  {E.}~\bibnamefont {Ostby}}, \bibinfo {author} {\bibfnamefont {K.~J.}\
  \bibnamefont {Vahala}}, \ and\ \bibinfo {author} {\bibfnamefont {H.~J.}\
  \bibnamefont {Kimble}},\ }\bibfield  {title} {\enquote {\bibinfo {title}
  {Strong interactions of single atoms and photons near a dielectric
  boundary},}\ }\href {\doibase 10.1038/nphys1837} {\bibfield  {journal}
  {\bibinfo  {journal} {Nature Physics}\ }\textbf {\bibinfo {volume} {7}},\
  \bibinfo {pages} {159--165} (\bibinfo {year} {2011})}\BibitemShut {NoStop}%
\bibitem [{\citenamefont {Volz}\ \emph {et~al.}(2014)\citenamefont {Volz},
  \citenamefont {Scheucher}, \citenamefont {Junge},\ and\ \citenamefont
  {Rauschenbeutel}}]{Volz2014}%
  \BibitemOpen
  \bibfield  {author} {\bibinfo {author} {\bibfnamefont {J.}~\bibnamefont
  {Volz}}, \bibinfo {author} {\bibfnamefont {M.}~\bibnamefont {Scheucher}},
  \bibinfo {author} {\bibfnamefont {C.}~\bibnamefont {Junge}}, \ and\ \bibinfo
  {author} {\bibfnamefont {A.}~\bibnamefont {Rauschenbeutel}},\ }\bibfield
  {title} {\enquote {\bibinfo {title} {Nonlinear $\pi$ phase shift for single
  fibre-guided photons interacting with a single resonator-enhanced atom},}\
  }\href@noop {} {\bibfield  {journal} {\bibinfo  {journal} {Nature Photonics}\
  }\textbf {\bibinfo {volume} {8}},\ \bibinfo {pages} {965} (\bibinfo {year}
  {2014})}\BibitemShut {NoStop}%
\bibitem [{\citenamefont {Goban}\ \emph {et~al.}(2014)\citenamefont {Goban},
  \citenamefont {Hung}, \citenamefont {Yu}, \citenamefont {Hood}, \citenamefont
  {Muniz}, \citenamefont {Lee}, \citenamefont {Martin}, \citenamefont
  {McClung}, \citenamefont {Choi}, \citenamefont {Chang}, \citenamefont
  {Painter},\ and\ \citenamefont {Kimble}}]{Goban2014}%
  \BibitemOpen
  \bibfield  {author} {\bibinfo {author} {\bibfnamefont {A.}~\bibnamefont
  {Goban}}, \bibinfo {author} {\bibfnamefont {C.-L.}\ \bibnamefont {Hung}},
  \bibinfo {author} {\bibfnamefont {S.-P.}\ \bibnamefont {Yu}}, \bibinfo
  {author} {\bibfnamefont {J.~D.}\ \bibnamefont {Hood}}, \bibinfo {author}
  {\bibfnamefont {J.~A.}\ \bibnamefont {Muniz}}, \bibinfo {author}
  {\bibfnamefont {J.~H.}\ \bibnamefont {Lee}}, \bibinfo {author} {\bibfnamefont
  {M.~J.}\ \bibnamefont {Martin}}, \bibinfo {author} {\bibfnamefont {A.~C.}\
  \bibnamefont {McClung}}, \bibinfo {author} {\bibfnamefont {K.~S.}\
  \bibnamefont {Choi}}, \bibinfo {author} {\bibfnamefont {D.~E.}\ \bibnamefont
  {Chang}}, \bibinfo {author} {\bibfnamefont {O.}~\bibnamefont {Painter}}, \
  and\ \bibinfo {author} {\bibfnamefont {H.~J.}\ \bibnamefont {Kimble}},\
  }\bibfield  {title} {\enquote {\bibinfo {title} {Atom–light interactions in
  photonic crystals},}\ }\href {\doibase 10.1038/ncomms4808} {\bibfield
  {journal} {\bibinfo  {journal} {Nature Communications}\ }\textbf {\bibinfo
  {volume} {5}},\ \bibinfo {pages} {3808} (\bibinfo {year} {2014})}\BibitemShut
  {NoStop}%
\bibitem [{\citenamefont {Goban}\ \emph {et~al.}(2015)\citenamefont {Goban},
  \citenamefont {Hung}, \citenamefont {Hood}, \citenamefont {Yu}, \citenamefont
  {Muniz}, \citenamefont {Painter},\ and\ \citenamefont {Kimble}}]{Goban2015}%
  \BibitemOpen
  \bibfield  {author} {\bibinfo {author} {\bibfnamefont {A.}~\bibnamefont
  {Goban}}, \bibinfo {author} {\bibfnamefont {C.~L.}\ \bibnamefont {Hung}},
  \bibinfo {author} {\bibfnamefont {J.}~\bibnamefont {Hood}}, \bibinfo {author}
  {\bibfnamefont {S.~P.}\ \bibnamefont {Yu}}, \bibinfo {author} {\bibfnamefont
  {J.}~\bibnamefont {Muniz}}, \bibinfo {author} {\bibfnamefont
  {O.}~\bibnamefont {Painter}}, \ and\ \bibinfo {author} {\bibfnamefont
  {H.}~\bibnamefont {Kimble}},\ }\bibfield  {title} {\enquote {\bibinfo {title}
  {Superradiance for atoms trapped along a photonic crystal waveguide},}\
  }\href {\doibase 10.1103/PhysRevLett.115.063601} {\bibfield  {journal}
  {\bibinfo  {journal} {Physical Review Letters}\ }\textbf {\bibinfo {volume}
  {115}},\ \bibinfo {pages} {063601} (\bibinfo {year} {2015})}\BibitemShut
  {NoStop}%
\bibitem [{\citenamefont {Hood}\ \emph {et~al.}(2016)\citenamefont {Hood},
  \citenamefont {Goban}, \citenamefont {Asenjo-Garcia}, \citenamefont {Lu},
  \citenamefont {Yu}, \citenamefont {Chang},\ and\ \citenamefont
  {Kimble}}]{Hood2016}%
  \BibitemOpen
  \bibfield  {author} {\bibinfo {author} {\bibfnamefont {J.~D.}\ \bibnamefont
  {Hood}}, \bibinfo {author} {\bibfnamefont {A.}~\bibnamefont {Goban}},
  \bibinfo {author} {\bibfnamefont {A.}~\bibnamefont {Asenjo-Garcia}}, \bibinfo
  {author} {\bibfnamefont {M.}~\bibnamefont {Lu}}, \bibinfo {author}
  {\bibfnamefont {S.~â.}\ \bibnamefont {Yu}}, \bibinfo {author} {\bibfnamefont
  {D.~E.}\ \bibnamefont {Chang}}, \ and\ \bibinfo {author} {\bibfnamefont
  {H.~J.}\ \bibnamefont {Kimble}},\ }\bibfield  {title} {\enquote {\bibinfo
  {title} {Atom–atom interactions around the band edge of a photonic crystal
  waveguide},}\ }\href {\doibase 10.1073/pnas.1603788113} {\bibfield  {journal}
  {\bibinfo  {journal} {Proccedings to National Academy of Science}\ }\textbf
  {\bibinfo {volume} {113}},\ \bibinfo {pages} {10507--10512} (\bibinfo {year}
  {2016})}\BibitemShut {NoStop}%
\bibitem [{\citenamefont {Englund}\ \emph {et~al.}(2008)\citenamefont
  {Englund}, \citenamefont {Fushman}, \citenamefont {Faraon},\ and\
  \citenamefont {Vuckovic}}]{Englund2008}%
  \BibitemOpen
  \bibfield  {author} {\bibinfo {author} {\bibfnamefont {D.}~\bibnamefont
  {Englund}}, \bibinfo {author} {\bibfnamefont {I.}~\bibnamefont {Fushman}},
  \bibinfo {author} {\bibfnamefont {A.}~\bibnamefont {Faraon}}, \ and\ \bibinfo
  {author} {\bibfnamefont {J.}~\bibnamefont {Vuckovic}},\ }\bibfield  {title}
  {\enquote {\bibinfo {title} {Quantum dots in photonic crystals: From quantum
  information processing to single photon nonlinear optics},}\ }\href {\doibase
  10.1016/j.photonics.2008.11.008} {\bibfield  {journal} {\bibinfo  {journal}
  {Photonics and Nanostructures Fundamentals and Applications}\ }\textbf
  {\bibinfo {volume} {7}},\ \bibinfo {pages} {56--62} (\bibinfo {year}
  {2008})}\BibitemShut {NoStop}%
\bibitem [{\citenamefont {Lodahl}, \citenamefont {Mahmoodian},\ and\
  \citenamefont {Stobbe}(2014)}]{Lodahl2014}%
  \BibitemOpen
  \bibfield  {author} {\bibinfo {author} {\bibfnamefont {P.}~\bibnamefont
  {Lodahl}}, \bibinfo {author} {\bibfnamefont {S.}~\bibnamefont {Mahmoodian}},
  \ and\ \bibinfo {author} {\bibfnamefont {S.}~\bibnamefont {Stobbe}},\
  }\bibfield  {title} {\enquote {\bibinfo {title} {Interfacing single photons
  and single quantum dots with photonic nanostructures},}\ }\href {\doibase
  10.1103/RevModPhys.87.347} {\bibfield  {journal} {\bibinfo  {journal} {Review
  of Modern Physics}\ }\textbf {\bibinfo {volume} {87}},\ \bibinfo {pages}
  {347--400} (\bibinfo {year} {2014})}\BibitemShut {NoStop}%
\bibitem [{\citenamefont {Khitrova}\ and\ \citenamefont
  {Gibbs}(2007)}]{Khitrova2007}%
  \BibitemOpen
  \bibfield  {author} {\bibinfo {author} {\bibfnamefont {G.}~\bibnamefont
  {Khitrova}}\ and\ \bibinfo {author} {\bibfnamefont {H.~M.}\ \bibnamefont
  {Gibbs}},\ }\bibfield  {title} {\enquote {\bibinfo {title} {Quantum dots:
  Collective radiance},}\ }\href@noop {} {\bibfield  {journal} {\bibinfo
  {journal} {Nature Physics}\ }\textbf {\bibinfo {volume} {3}},\ \bibinfo
  {pages} {84} (\bibinfo {year} {2007})}\BibitemShut {NoStop}%
\bibitem [{\citenamefont {Patel}\ \emph {et~al.}(2016)\citenamefont {Patel},
  \citenamefont {Schröder}, \citenamefont {W.}, \citenamefont {Li},
  \citenamefont {Mouradian}, \citenamefont {Chen},\ and\ \citenamefont
  {Englund}}]{Patel2016}%
  \BibitemOpen
  \bibfield  {author} {\bibinfo {author} {\bibfnamefont {R.~N.}\ \bibnamefont
  {Patel}}, \bibinfo {author} {\bibfnamefont {T.}~\bibnamefont {Schröder}},
  \bibinfo {author} {\bibfnamefont {N.}~\bibnamefont {W.}}, \bibinfo {author}
  {\bibfnamefont {L.}~\bibnamefont {Li}}, \bibinfo {author} {\bibfnamefont
  {S.~L.}\ \bibnamefont {Mouradian}}, \bibinfo {author} {\bibfnamefont {E.~H.}\
  \bibnamefont {Chen}}, \ and\ \bibinfo {author} {\bibfnamefont {D.~R.}\
  \bibnamefont {Englund}},\ }\bibfield  {title} {\enquote {\bibinfo {title}
  {Efficient photon coupling from a diamond nitrogen vacancy center by
  integration with silica fiber},}\ }\href {\doibase 10.1038/lsa.2016.32}
  {\bibfield  {journal} {\bibinfo  {journal} {Light: Science and Applications}\
  }\textbf {\bibinfo {volume} {5}},\ \bibinfo {pages} {e16032} (\bibinfo {year}
  {2016})}\BibitemShut {NoStop}%
\bibitem [{\citenamefont {Sipahigil}\ \emph {et~al.}(2016)\citenamefont
  {Sipahigil}, \citenamefont {Evans}, \citenamefont {Sukachev}, \citenamefont
  {Burek}, \citenamefont {Borregaard}, \citenamefont {Bhaskar}, \citenamefont
  {Nguyen}, \citenamefont {Pacheco}, \citenamefont {Atikian}, \citenamefont
  {Meuwly}, \citenamefont {Camacho}, \citenamefont {Jelezko}, \citenamefont
  {Bielejec}, \citenamefont {Park}, \citenamefont {Lon{\v c}ar},\ and\
  \citenamefont {Lukin}}]{Sipahigil2016}%
  \BibitemOpen
  \bibfield  {author} {\bibinfo {author} {\bibfnamefont {A.}~\bibnamefont
  {Sipahigil}}, \bibinfo {author} {\bibfnamefont {R.~E.}\ \bibnamefont
  {Evans}}, \bibinfo {author} {\bibfnamefont {D.~D.}\ \bibnamefont {Sukachev}},
  \bibinfo {author} {\bibfnamefont {M.~J.}\ \bibnamefont {Burek}}, \bibinfo
  {author} {\bibfnamefont {J.}~\bibnamefont {Borregaard}}, \bibinfo {author}
  {\bibfnamefont {M.~K.}\ \bibnamefont {Bhaskar}}, \bibinfo {author}
  {\bibfnamefont {C.~T.}\ \bibnamefont {Nguyen}}, \bibinfo {author}
  {\bibfnamefont {J.~L.}\ \bibnamefont {Pacheco}}, \bibinfo {author}
  {\bibfnamefont {H.~A.}\ \bibnamefont {Atikian}}, \bibinfo {author}
  {\bibfnamefont {C.}~\bibnamefont {Meuwly}}, \bibinfo {author} {\bibfnamefont
  {R.~M.}\ \bibnamefont {Camacho}}, \bibinfo {author} {\bibfnamefont
  {F.}~\bibnamefont {Jelezko}}, \bibinfo {author} {\bibfnamefont
  {E.}~\bibnamefont {Bielejec}}, \bibinfo {author} {\bibfnamefont
  {H.}~\bibnamefont {Park}}, \bibinfo {author} {\bibfnamefont {M.}~\bibnamefont
  {Lon{\v c}ar}}, \ and\ \bibinfo {author} {\bibfnamefont {M.~D.}\ \bibnamefont
  {Lukin}},\ }\bibfield  {title} {\enquote {\bibinfo {title} {An integrated
  diamond nanophotonics platform for quantum optical networks},}\ }\href
  {\doibase 10.1126/science.aah6875} {\bibfield  {journal} {\bibinfo  {journal}
  {Science}\ } (\bibinfo {year} {2016}),\ 10.1126/science.aah6875}\BibitemShut
  {NoStop}%
\bibitem [{\citenamefont {G\"undo\ifmmode~\breve{g}\else \u{g}\fi{}an}\ \emph
  {et~al.}(2015)\citenamefont {G\"undo\ifmmode~\breve{g}\else \u{g}\fi{}an},
  \citenamefont {Ledingham}, \citenamefont {Kutluer}, \citenamefont {Mazzera},\
  and\ \citenamefont {de~Riedmatten}}]{Riedmatten2015}%
  \BibitemOpen
  \bibfield  {author} {\bibinfo {author} {\bibfnamefont {M.}~\bibnamefont
  {G\"undo\ifmmode~\breve{g}\else \u{g}\fi{}an}}, \bibinfo {author}
  {\bibfnamefont {P.~M.}\ \bibnamefont {Ledingham}}, \bibinfo {author}
  {\bibfnamefont {K.}~\bibnamefont {Kutluer}}, \bibinfo {author} {\bibfnamefont
  {M.}~\bibnamefont {Mazzera}}, \ and\ \bibinfo {author} {\bibfnamefont
  {H.}~\bibnamefont {de~Riedmatten}},\ }\bibfield  {title} {\enquote {\bibinfo
  {title} {Solid state spin-wave quantum memory for time-bin qubits},}\ }\href
  {\doibase 10.1103/PhysRevLett.114.230501} {\bibfield  {journal} {\bibinfo
  {journal} {Phys. Rev. Lett.}\ }\textbf {\bibinfo {volume} {114}},\ \bibinfo
  {pages} {230501} (\bibinfo {year} {2015})}\BibitemShut {NoStop}%
\bibitem [{\citenamefont {Zhong}\ \emph {et~al.}(2015)\citenamefont {Zhong},
  \citenamefont {Kindem}, \citenamefont {Miyazono},\ and\ \citenamefont
  {Faraon}}]{Zhong2015}%
  \BibitemOpen
  \bibfield  {author} {\bibinfo {author} {\bibfnamefont {T.}~\bibnamefont
  {Zhong}}, \bibinfo {author} {\bibfnamefont {J.~M.}\ \bibnamefont {Kindem}},
  \bibinfo {author} {\bibfnamefont {E.}~\bibnamefont {Miyazono}}, \ and\
  \bibinfo {author} {\bibfnamefont {A.}~\bibnamefont {Faraon}},\ }\bibfield
  {title} {\enquote {\bibinfo {title} {Nanophotonic coherent light–matter
  interfaces based on rare-earth-doped crystals},}\ }\href {\doibase
  10.1038/ncomms9206} {\bibfield  {journal} {\bibinfo  {journal} {Nature
  Communications}\ }\textbf {\bibinfo {volume} {6}},\ \bibinfo {pages} {8206}
  (\bibinfo {year} {2015})}\BibitemShut {NoStop}%
\bibitem [{\citenamefont {Dibos}\ \emph {et~al.}(2018)\citenamefont {Dibos},
  \citenamefont {Raha}, \citenamefont {Phenicie},\ and\ \citenamefont
  {Thompson}}]{Dibos2017}%
  \BibitemOpen
  \bibfield  {author} {\bibinfo {author} {\bibfnamefont {A.~M.}\ \bibnamefont
  {Dibos}}, \bibinfo {author} {\bibfnamefont {M.}~\bibnamefont {Raha}},
  \bibinfo {author} {\bibfnamefont {C.~M.}\ \bibnamefont {Phenicie}}, \ and\
  \bibinfo {author} {\bibfnamefont {J.~D.}\ \bibnamefont {Thompson}},\
  }\bibfield  {title} {\enquote {\bibinfo {title} {Atomic source of single
  photons in the telecom band},}\ }\href {\doibase
  10.1103/PhysRevLett.120.243601} {\bibfield  {journal} {\bibinfo  {journal}
  {Phys. Rev. Lett.}\ }\textbf {\bibinfo {volume} {120}},\ \bibinfo {pages}
  {243601} (\bibinfo {year} {2018})}\BibitemShut {NoStop}%
\bibitem [{\citenamefont {Simon}\ \emph {et~al.}(2011)\citenamefont {Simon},
  \citenamefont {Bakr}, \citenamefont {Ma}, \citenamefont {Tai}, \citenamefont
  {Preiss},\ and\ \citenamefont {Greiner}}]{Simon2011}%
  \BibitemOpen
  \bibfield  {author} {\bibinfo {author} {\bibfnamefont {J.}~\bibnamefont
  {Simon}}, \bibinfo {author} {\bibfnamefont {W.~S.}\ \bibnamefont {Bakr}},
  \bibinfo {author} {\bibfnamefont {R.}~\bibnamefont {Ma}}, \bibinfo {author}
  {\bibfnamefont {M.~E.}\ \bibnamefont {Tai}}, \bibinfo {author} {\bibfnamefont
  {P.~M.}\ \bibnamefont {Preiss}}, \ and\ \bibinfo {author} {\bibfnamefont
  {M.}~\bibnamefont {Greiner}},\ }\bibfield  {title} {\enquote {\bibinfo
  {title} {Quantum simulation of antiferromagnetic spin chains in an optical
  lattice},}\ }\href {\doibase 10.1038/nature09994} {\bibfield  {journal}
  {\bibinfo  {journal} {Nature}\ }\textbf {\bibinfo {volume} {472}},\ \bibinfo
  {pages} {307--312} (\bibinfo {year} {2011})}\BibitemShut {NoStop}%
\bibitem [{\citenamefont {Uehlinger}\ \emph {et~al.}(2013)\citenamefont
  {Uehlinger}, \citenamefont {Jotzu}, \citenamefont {Messer}, \citenamefont
  {Greif}, \citenamefont {Hofstetter}, \citenamefont {Bissbort},\ and\
  \citenamefont {Esslinger}}]{Uehlinger2013}%
  \BibitemOpen
  \bibfield  {author} {\bibinfo {author} {\bibfnamefont {T.}~\bibnamefont
  {Uehlinger}}, \bibinfo {author} {\bibfnamefont {G.}~\bibnamefont {Jotzu}},
  \bibinfo {author} {\bibfnamefont {M.}~\bibnamefont {Messer}}, \bibinfo
  {author} {\bibfnamefont {D.}~\bibnamefont {Greif}}, \bibinfo {author}
  {\bibfnamefont {W.}~\bibnamefont {Hofstetter}}, \bibinfo {author}
  {\bibfnamefont {U.}~\bibnamefont {Bissbort}}, \ and\ \bibinfo {author}
  {\bibfnamefont {T.}~\bibnamefont {Esslinger}},\ }\bibfield  {title} {\enquote
  {\bibinfo {title} {Artificial graphene with tunable interactions},}\ }\href
  {\doibase 10.1103/PhysRevLett.111.185307} {\bibfield  {journal} {\bibinfo
  {journal} {Physical Review Letters}\ }\textbf {\bibinfo {volume} {111}},\
  \bibinfo {pages} {185307} (\bibinfo {year} {2013})}\BibitemShut {NoStop}%
\bibitem [{\citenamefont {Gonzalez-Tudela}\ \emph {et~al.}(2015)\citenamefont
  {Gonzalez-Tudela}, \citenamefont {Hung}, \citenamefont {Chang}, \citenamefont
  {Cirac},\ and\ \citenamefont {Kimble}}]{Tudela2015}%
  \BibitemOpen
  \bibfield  {author} {\bibinfo {author} {\bibfnamefont {A.}~\bibnamefont
  {Gonzalez-Tudela}}, \bibinfo {author} {\bibfnamefont {C.-L.}\ \bibnamefont
  {Hung}}, \bibinfo {author} {\bibfnamefont {D.~E.}\ \bibnamefont {Chang}},
  \bibinfo {author} {\bibfnamefont {J.~I.}\ \bibnamefont {Cirac}}, \ and\
  \bibinfo {author} {\bibfnamefont {H.~J.}\ \bibnamefont {Kimble}},\ }\bibfield
   {title} {\enquote {\bibinfo {title} {Subwavelength vacuum lattices and
  atom–atom interactions in two-dimensional photonic crystals},}\ }\href
  {\doibase 10.1038/NPHOTON.2015.54} {\bibfield  {journal} {\bibinfo  {journal}
  {Nature Photonics}\ }\textbf {\bibinfo {volume} {9}},\ \bibinfo {pages}
  {320--325} (\bibinfo {year} {2015})}\BibitemShut {NoStop}%
\bibitem [{\citenamefont {Kollar}\ \emph {et~al.}(2017)\citenamefont {Kollar},
  \citenamefont {Papageorge}, \citenamefont {Vaidya}, \citenamefont {Guo},
  \citenamefont {Keeling},\ and\ \citenamefont {Lev}}]{Kollar2017}%
  \BibitemOpen
  \bibfield  {author} {\bibinfo {author} {\bibfnamefont {A.~J.}\ \bibnamefont
  {Kollar}}, \bibinfo {author} {\bibfnamefont {A.~T.}\ \bibnamefont
  {Papageorge}}, \bibinfo {author} {\bibfnamefont {V.~D.}\ \bibnamefont
  {Vaidya}}, \bibinfo {author} {\bibfnamefont {Y.}~\bibnamefont {Guo}},
  \bibinfo {author} {\bibfnamefont {J.}~\bibnamefont {Keeling}}, \ and\
  \bibinfo {author} {\bibfnamefont {B.~L.}\ \bibnamefont {Lev}},\ }\bibfield
  {title} {\enquote {\bibinfo {title} {Supermode-density-wave-polariton
  condensation with a bose–einstein condensate in a multimode cavity},}\
  }\href {\doibase 10.1038/ncomms14386} {\bibfield  {journal} {\bibinfo
  {journal} {Nature Communications}\ }\textbf {\bibinfo {volume} {8}},\
  \bibinfo {pages} {14386} (\bibinfo {year} {2017})}\BibitemShut {NoStop}%
\bibitem [{\citenamefont {Perczel}\ \emph {et~al.}(2017)\citenamefont
  {Perczel}, \citenamefont {Borregaard}, \citenamefont {Chang}, \citenamefont
  {Pichler}, \citenamefont {Yelin}, \citenamefont {Zoller},\ and\ \citenamefont
  {Lukin}}]{Perczel_PRL2017}%
  \BibitemOpen
  \bibfield  {author} {\bibinfo {author} {\bibfnamefont {J.}~\bibnamefont
  {Perczel}}, \bibinfo {author} {\bibfnamefont {J.}~\bibnamefont {Borregaard}},
  \bibinfo {author} {\bibfnamefont {D.~E.}\ \bibnamefont {Chang}}, \bibinfo
  {author} {\bibfnamefont {H.}~\bibnamefont {Pichler}}, \bibinfo {author}
  {\bibfnamefont {S.~F.}\ \bibnamefont {Yelin}}, \bibinfo {author}
  {\bibfnamefont {P.}~\bibnamefont {Zoller}}, \ and\ \bibinfo {author}
  {\bibfnamefont {M.~D.}\ \bibnamefont {Lukin}},\ }\bibfield  {title} {\enquote
  {\bibinfo {title} {Topological quantum optics in two-dimensional atomic
  arrays},}\ }\href {\doibase 10.1103/PhysRevLett.119.023603} {\bibfield
  {journal} {\bibinfo  {journal} {Phys. Rev. Lett.}\ }\textbf {\bibinfo
  {volume} {119}},\ \bibinfo {pages} {023603} (\bibinfo {year}
  {2017})}\BibitemShut {NoStop}%
\bibitem [{\citenamefont {Shi}, \citenamefont {Kimble},\ and\ \citenamefont
  {Cirac}(2017)}]{Shi2017}%
  \BibitemOpen
  \bibfield  {author} {\bibinfo {author} {\bibfnamefont {T.}~\bibnamefont
  {Shi}}, \bibinfo {author} {\bibfnamefont {H.}~\bibnamefont {Kimble}}, \ and\
  \bibinfo {author} {\bibfnamefont {J.}~\bibnamefont {Cirac}},\ }\bibfield
  {title} {\enquote {\bibinfo {title} {Topological phenomena in classical
  optical networks},}\ }\href@noop {} {\bibfield  {journal} {\bibinfo
  {journal} {Proceedings of the National Academy of Sciences}\ ,\ \bibinfo
  {pages} {201708944}} (\bibinfo {year} {2017})}\BibitemShut {NoStop}%
\bibitem [{\citenamefont {Bloch}(2005)}]{Bloch2005}%
  \BibitemOpen
  \bibfield  {author} {\bibinfo {author} {\bibfnamefont {I.}~\bibnamefont
  {Bloch}},\ }\bibfield  {title} {\enquote {\bibinfo {title} {Ultracold quantum
  gases in optical lattices},}\ }\href@noop {} {\bibfield  {journal} {\bibinfo
  {journal} {Nature Physics}\ }\textbf {\bibinfo {volume} {1}},\ \bibinfo
  {pages} {23} (\bibinfo {year} {2005})}\BibitemShut {NoStop}%
\bibitem [{\citenamefont {Becker}\ \emph {et~al.}(2010)\citenamefont {Becker},
  \citenamefont {Soltan-Panahi}, \citenamefont {Kronjager}, \citenamefont
  {Dorscher}, \citenamefont {Bongs},\ and\ \citenamefont
  {Sengstock}}]{Becker2010}%
  \BibitemOpen
  \bibfield  {author} {\bibinfo {author} {\bibfnamefont {C.}~\bibnamefont
  {Becker}}, \bibinfo {author} {\bibfnamefont {P.}~\bibnamefont
  {Soltan-Panahi}}, \bibinfo {author} {\bibfnamefont {J.}~\bibnamefont
  {Kronjager}}, \bibinfo {author} {\bibfnamefont {S.}~\bibnamefont {Dorscher}},
  \bibinfo {author} {\bibfnamefont {K.}~\bibnamefont {Bongs}}, \ and\ \bibinfo
  {author} {\bibfnamefont {K.}~\bibnamefont {Sengstock}},\ }\bibfield  {title}
  {\enquote {\bibinfo {title} {Ultracold quantum gases in triangular optical
  lattices},}\ }\href@noop {} {\bibfield  {journal} {\bibinfo  {journal} {New
  Journal of Physics}\ }\textbf {\bibinfo {volume} {12}},\ \bibinfo {pages}
  {065025} (\bibinfo {year} {2010})}\BibitemShut {NoStop}%
\bibitem [{\citenamefont {Mazurenko}\ \emph {et~al.}(2017)\citenamefont
  {Mazurenko}, \citenamefont {Chiu}, \citenamefont {Ji}, \citenamefont
  {Parsons}, \citenamefont {Kanasz-Nagy}, \citenamefont {Schmidt},
  \citenamefont {Grusdt}, \citenamefont {Demler}, \citenamefont {Greif},\ and\
  \citenamefont {Greiner}}]{Mazurenko2017}%
  \BibitemOpen
  \bibfield  {author} {\bibinfo {author} {\bibfnamefont {A.}~\bibnamefont
  {Mazurenko}}, \bibinfo {author} {\bibfnamefont {C.~S.}\ \bibnamefont {Chiu}},
  \bibinfo {author} {\bibfnamefont {G.}~\bibnamefont {Ji}}, \bibinfo {author}
  {\bibfnamefont {M.~F.}\ \bibnamefont {Parsons}}, \bibinfo {author}
  {\bibfnamefont {M.}~\bibnamefont {Kanasz-Nagy}}, \bibinfo {author}
  {\bibfnamefont {R.}~\bibnamefont {Schmidt}}, \bibinfo {author} {\bibfnamefont
  {F.}~\bibnamefont {Grusdt}}, \bibinfo {author} {\bibfnamefont
  {E.}~\bibnamefont {Demler}}, \bibinfo {author} {\bibfnamefont
  {D.}~\bibnamefont {Greif}}, \ and\ \bibinfo {author} {\bibfnamefont
  {M.}~\bibnamefont {Greiner}},\ }\bibfield  {title} {\enquote {\bibinfo
  {title} {A cold-atom fermi–hubbard antiferromagnet},}\ }\href {\doibase
  10.1038/nature22362} {\bibfield  {journal} {\bibinfo  {journal} {Nature}\
  }\textbf {\bibinfo {volume} {545}},\ \bibinfo {pages} {462--466} (\bibinfo
  {year} {2017})}\BibitemShut {NoStop}%
\bibitem [{\citenamefont {Jaksch}\ and\ \citenamefont
  {Zoller}(2005)}]{Jaksch2005}%
  \BibitemOpen
  \bibfield  {author} {\bibinfo {author} {\bibfnamefont {D.}~\bibnamefont
  {Jaksch}}\ and\ \bibinfo {author} {\bibfnamefont {P.}~\bibnamefont
  {Zoller}},\ }\bibfield  {title} {\enquote {\bibinfo {title} {The cold atom
  hubbard toolbox},}\ }\href {\doibase 10.1016/j.aop.2004.09.010} {\bibfield
  {journal} {\bibinfo  {journal} {Annals of Physics}\ }\textbf {\bibinfo
  {volume} {315}},\ \bibinfo {pages} {52--79} (\bibinfo {year}
  {2005})}\BibitemShut {NoStop}%
\bibitem [{\citenamefont {Weitenberg}\ \emph {et~al.}(2011)\citenamefont
  {Weitenberg}, \citenamefont {Endres}, \citenamefont {Sherson}, \citenamefont
  {Cheneau}, \citenamefont {Schauß}, \citenamefont {Fukuhara}, \citenamefont
  {Bloch},\ and\ \citenamefont {Kuhr}}]{Weitenberg2011}%
  \BibitemOpen
  \bibfield  {author} {\bibinfo {author} {\bibfnamefont {C.}~\bibnamefont
  {Weitenberg}}, \bibinfo {author} {\bibfnamefont {M.}~\bibnamefont {Endres}},
  \bibinfo {author} {\bibfnamefont {J.~F.}\ \bibnamefont {Sherson}}, \bibinfo
  {author} {\bibfnamefont {M.}~\bibnamefont {Cheneau}}, \bibinfo {author}
  {\bibfnamefont {P.}~\bibnamefont {Schauß}}, \bibinfo {author} {\bibfnamefont
  {T.}~\bibnamefont {Fukuhara}}, \bibinfo {author} {\bibfnamefont
  {I.}~\bibnamefont {Bloch}}, \ and\ \bibinfo {author} {\bibfnamefont
  {S.}~\bibnamefont {Kuhr}},\ }\bibfield  {title} {\enquote {\bibinfo {title}
  {Single-spin addressing in an atomic mott insulator},}\ }\href {\doibase
  10.1038/nature09827} {\bibfield  {journal} {\bibinfo  {journal} {Nature}\
  }\textbf {\bibinfo {volume} {471}},\ \bibinfo {pages} {319--324} (\bibinfo
  {year} {2011})}\BibitemShut {NoStop}%
\bibitem [{\citenamefont {Struck}\ \emph {et~al.}(2011)\citenamefont {Struck},
  \citenamefont {Olschlager}, \citenamefont {Targat}, \citenamefont
  {Soltan-Panahi}, \citenamefont {Eckardt}, \citenamefont {Lewenstein},
  \citenamefont {Windpassinger},\ and\ \citenamefont {Sengstock}}]{Struck2011}%
  \BibitemOpen
  \bibfield  {author} {\bibinfo {author} {\bibfnamefont {J.}~\bibnamefont
  {Struck}}, \bibinfo {author} {\bibfnamefont {C.}~\bibnamefont {Olschlager}},
  \bibinfo {author} {\bibfnamefont {R.~L.}\ \bibnamefont {Targat}}, \bibinfo
  {author} {\bibfnamefont {P.}~\bibnamefont {Soltan-Panahi}}, \bibinfo {author}
  {\bibfnamefont {A.}~\bibnamefont {Eckardt}}, \bibinfo {author} {\bibfnamefont
  {M.}~\bibnamefont {Lewenstein}}, \bibinfo {author} {\bibfnamefont
  {P.}~\bibnamefont {Windpassinger}}, \ and\ \bibinfo {author} {\bibfnamefont
  {K.}~\bibnamefont {Sengstock}},\ }\bibfield  {title} {\enquote {\bibinfo
  {title} {Quantum simulation of frustrated classical magnetism in triangular
  optical lattices},}\ }\href {\doibase 10.1126/science.1207239} {\bibfield
  {journal} {\bibinfo  {journal} {Science}\ }\textbf {\bibinfo {volume}
  {333}},\ \bibinfo {pages} {996--999} (\bibinfo {year} {2011})}\BibitemShut
  {NoStop}%
\bibitem [{\citenamefont {Gonz\'alez-Tudela}\ and\ \citenamefont
  {Cirac}(2017{\natexlab{a}})}]{Tudela_PRL2017}%
  \BibitemOpen
  \bibfield  {author} {\bibinfo {author} {\bibfnamefont {A.}~\bibnamefont
  {Gonz\'alez-Tudela}}\ and\ \bibinfo {author} {\bibfnamefont {J.~I.}\
  \bibnamefont {Cirac}},\ }\bibfield  {title} {\enquote {\bibinfo {title}
  {Quantum emitters in two-dimensional structured reservoirs in the
  nonperturbative regime},}\ }\href {\doibase 10.1103/PhysRevLett.119.143602}
  {\bibfield  {journal} {\bibinfo  {journal} {Phys. Rev. Lett.}\ }\textbf
  {\bibinfo {volume} {119}},\ \bibinfo {pages} {143602} (\bibinfo {year}
  {2017}{\natexlab{a}})}\BibitemShut {NoStop}%
\bibitem [{\citenamefont {Gonz\'alez-Tudela}\ and\ \citenamefont
  {Cirac}(2017{\natexlab{b}})}]{Tudela_PRA2017}%
  \BibitemOpen
  \bibfield  {author} {\bibinfo {author} {\bibfnamefont {A.}~\bibnamefont
  {Gonz\'alez-Tudela}}\ and\ \bibinfo {author} {\bibfnamefont {J.~I.}\
  \bibnamefont {Cirac}},\ }\bibfield  {title} {\enquote {\bibinfo {title}
  {Markovian and non-markovian dynamics of quantum emitters coupled to
  two-dimensional structured reservoirs},}\ }\href {\doibase
  10.1103/PhysRevA.96.043811} {\bibfield  {journal} {\bibinfo  {journal} {Phys.
  Rev. A}\ }\textbf {\bibinfo {volume} {96}},\ \bibinfo {pages} {043811}
  (\bibinfo {year} {2017}{\natexlab{b}})}\BibitemShut {NoStop}%
\bibitem [{\citenamefont {Muniz}(2017)}]{MunizPhD}%
  \BibitemOpen
  \bibfield  {author} {\bibinfo {author} {\bibfnamefont {J.~A.}\ \bibnamefont
  {Muniz}},\ }\emph {\bibinfo {title} {Nanoscopic Atomic Lattices with
  Light-Mediated Interactions}},\ \href@noop {} {Ph.D. thesis},\ \bibinfo
  {school} {Caltech} (\bibinfo {year} {2017})\BibitemShut {NoStop}%
\bibitem [{\citenamefont {Yu}(2017)}]{YuPhD}%
  \BibitemOpen
  \bibfield  {author} {\bibinfo {author} {\bibfnamefont {S.-P.}\ \bibnamefont
  {Yu}},\ }\emph {\bibinfo {title} {Nano-Photonic Platform for Atom-Light
  Interaction}},\ \href@noop {} {Ph.D. thesis},\ \bibinfo  {school} {Caltech}
  (\bibinfo {year} {2017})\BibitemShut {NoStop}%
\bibitem [{\citenamefont {Yu}\ \emph {et~al.}(2014)\citenamefont {Yu},
  \citenamefont {Hood}, \citenamefont {Muniz}, \citenamefont {Martin},
  \citenamefont {Norte}, \citenamefont {Hung}, \citenamefont {Meenehan},
  \citenamefont {Cohen}, \citenamefont {Painter},\ and\ \citenamefont
  {Kimble}}]{Yu2014}%
  \BibitemOpen
  \bibfield  {author} {\bibinfo {author} {\bibfnamefont {S.-P.}\ \bibnamefont
  {Yu}}, \bibinfo {author} {\bibfnamefont {J.~D.}\ \bibnamefont {Hood}},
  \bibinfo {author} {\bibfnamefont {J.~A.}\ \bibnamefont {Muniz}}, \bibinfo
  {author} {\bibfnamefont {M.~J.}\ \bibnamefont {Martin}}, \bibinfo {author}
  {\bibfnamefont {R.}~\bibnamefont {Norte}}, \bibinfo {author} {\bibfnamefont
  {C.-L.}\ \bibnamefont {Hung}}, \bibinfo {author} {\bibfnamefont {S.~M.}\
  \bibnamefont {Meenehan}}, \bibinfo {author} {\bibfnamefont {J.~D.}\
  \bibnamefont {Cohen}}, \bibinfo {author} {\bibfnamefont {O.}~\bibnamefont
  {Painter}}, \ and\ \bibinfo {author} {\bibfnamefont {H.~J.}\ \bibnamefont
  {Kimble}},\ }\bibfield  {title} {\enquote {\bibinfo {title} {Nanowire
  photonic crystal waveguides for single-atom trapping and strong light-matter
  interactions},}\ }\href {\doibase 10.1063/1.4868975} {\bibfield  {journal}
  {\bibinfo  {journal} {Applied Physics Letters}\ }\textbf {\bibinfo {volume}
  {104}},\ \bibinfo {pages} {111103} (\bibinfo {year} {2014})}\BibitemShut
  {NoStop}%
\bibitem [{\citenamefont {Kosaka}\ \emph {et~al.}(1999)\citenamefont {Kosaka},
  \citenamefont {Kawashima}, \citenamefont {Tomita}, \citenamefont {Notomi},
  \citenamefont {Tamamura}, \citenamefont {Sato},\ and\ \citenamefont
  {Kawakami}}]{Kosaka1999}%
  \BibitemOpen
  \bibfield  {author} {\bibinfo {author} {\bibfnamefont {H.}~\bibnamefont
  {Kosaka}}, \bibinfo {author} {\bibfnamefont {T.}~\bibnamefont {Kawashima}},
  \bibinfo {author} {\bibfnamefont {A.}~\bibnamefont {Tomita}}, \bibinfo
  {author} {\bibfnamefont {M.}~\bibnamefont {Notomi}}, \bibinfo {author}
  {\bibfnamefont {T.}~\bibnamefont {Tamamura}}, \bibinfo {author}
  {\bibfnamefont {T.}~\bibnamefont {Sato}}, \ and\ \bibinfo {author}
  {\bibfnamefont {S.}~\bibnamefont {Kawakami}},\ }\bibfield  {title} {\enquote
  {\bibinfo {title} {Self-collimating phenomena in photonic crystals},}\ }\href
  {\doibase 10.1063/1.123502} {\bibfield  {journal} {\bibinfo  {journal}
  {Applied Physics Letters}\ }\textbf {\bibinfo {volume} {74}},\ \bibinfo
  {pages} {1212--1214} (\bibinfo {year} {1999})},\ \Eprint
  {http://arxiv.org/abs/https://doi.org/10.1063/1.123502}
  {https://doi.org/10.1063/1.123502} \BibitemShut {NoStop}%
\bibitem [{\citenamefont {Iliew}\ \emph {et~al.}(2004)\citenamefont {Iliew},
  \citenamefont {Etrich}, \citenamefont {Peschel}, \citenamefont {Lederer},
  \citenamefont {Augustin}, \citenamefont {Fuchs}, \citenamefont {Schelle},
  \citenamefont {Kley}, \citenamefont {Nolte},\ and\ \citenamefont
  {T{\"u}nnermann}}]{Iliewa2004}%
  \BibitemOpen
  \bibfield  {author} {\bibinfo {author} {\bibfnamefont {R.}~\bibnamefont
  {Iliew}}, \bibinfo {author} {\bibfnamefont {C.}~\bibnamefont {Etrich}},
  \bibinfo {author} {\bibfnamefont {U.}~\bibnamefont {Peschel}}, \bibinfo
  {author} {\bibfnamefont {F.}~\bibnamefont {Lederer}}, \bibinfo {author}
  {\bibfnamefont {M.}~\bibnamefont {Augustin}}, \bibinfo {author}
  {\bibfnamefont {H.-J.}\ \bibnamefont {Fuchs}}, \bibinfo {author}
  {\bibfnamefont {D.}~\bibnamefont {Schelle}}, \bibinfo {author} {\bibfnamefont
  {E.-B.}\ \bibnamefont {Kley}}, \bibinfo {author} {\bibfnamefont
  {S.}~\bibnamefont {Nolte}}, \ and\ \bibinfo {author} {\bibfnamefont
  {A.}~\bibnamefont {T{\"u}nnermann}},\ }\bibfield  {title} {\enquote {\bibinfo
  {title} {Diffractionless propagation of light in a low-index photonic-crystal
  film},}\ }\href@noop {} {\bibfield  {journal} {\bibinfo  {journal} {Applied
  physics letters}\ }\textbf {\bibinfo {volume} {85}},\ \bibinfo {pages}
  {5854--5856} (\bibinfo {year} {2004})}\BibitemShut {NoStop}%
\bibitem [{\citenamefont {Agarwal}(1975)}]{Agarwal1975}%
  \BibitemOpen
  \bibfield  {author} {\bibinfo {author} {\bibfnamefont {G.~S.}\ \bibnamefont
  {Agarwal}},\ }\bibfield  {title} {\enquote {\bibinfo {title} {Quantum
  electrodynamics in the presence of dielectrics and conductors. iv. general
  theory for spontaneous emission in finite geometries},}\ }\href {\doibase
  10.1103/PhysRevA.12.1475} {\bibfield  {journal} {\bibinfo  {journal}
  {Physical Review A}\ }\textbf {\bibinfo {volume} {12}},\ \bibinfo {pages}
  {1475} (\bibinfo {year} {1975})}\BibitemShut {NoStop}%
\bibitem [{\citenamefont {Novotny}\ and\ \citenamefont
  {Hecht}(2012)}]{Novotny2012}%
  \BibitemOpen
  \bibfield  {author} {\bibinfo {author} {\bibfnamefont {L.}~\bibnamefont
  {Novotny}}\ and\ \bibinfo {author} {\bibfnamefont {B.}~\bibnamefont
  {Hecht}},\ }\href@noop {} {\emph {\bibinfo {title} {Principles of
  Nano-optics}}}\ (\bibinfo  {publisher} {Cambridge university press},\
  \bibinfo {year} {2012})\BibitemShut {NoStop}%
\bibitem [{\citenamefont {Hung}\ \emph {et~al.}(2013)\citenamefont {Hung},
  \citenamefont {Meenehan}, \citenamefont {Chang}, \citenamefont {Painter},\
  and\ \citenamefont {Kimble}}]{Hung2013}%
  \BibitemOpen
  \bibfield  {author} {\bibinfo {author} {\bibfnamefont {C.-L.}\ \bibnamefont
  {Hung}}, \bibinfo {author} {\bibfnamefont {S.~M.}\ \bibnamefont {Meenehan}},
  \bibinfo {author} {\bibfnamefont {D.~E.}\ \bibnamefont {Chang}}, \bibinfo
  {author} {\bibfnamefont {O.}~\bibnamefont {Painter}}, \ and\ \bibinfo
  {author} {\bibfnamefont {H.~J.}\ \bibnamefont {Kimble}},\ }\bibfield  {title}
  {\enquote {\bibinfo {title} {Trapped atoms in one-dimensional photonic
  crystals},}\ }\href {\doibase 10.1088/1367-2630/15/8/08302} {\bibfield
  {journal} {\bibinfo  {journal} {New Journal of Physics}\ }\textbf {\bibinfo
  {volume} {15}},\ \bibinfo {pages} {083026} (\bibinfo {year}
  {2013})}\BibitemShut {NoStop}%
\bibitem [{\citenamefont {Steck}(2010)}]{Steck2010}%
  \BibitemOpen
  \bibfield  {author} {\bibinfo {author} {\bibfnamefont {D.~A.}\ \bibnamefont
  {Steck}},\ }\href {http://steck.us/alkalidata/} {\enquote {\bibinfo {title}
  {Cesium d line data},}\ } (\bibinfo {year} {2010})\BibitemShut {NoStop}%
\bibitem [{\citenamefont {Roy-Choudhury}\ and\ \citenamefont
  {Hughes}(2015)}]{Hughes2015}%
  \BibitemOpen
  \bibfield  {author} {\bibinfo {author} {\bibfnamefont {K.}~\bibnamefont
  {Roy-Choudhury}}\ and\ \bibinfo {author} {\bibfnamefont {S.}~\bibnamefont
  {Hughes}},\ }\bibfield  {title} {\enquote {\bibinfo {title} {Quantum theory
  of the emission spectrum from quantum dots coupled to structured photonic
  reservoirs and acoustic phonons},}\ }\href {\doibase
  10.1103/PhysRevB.92.205406} {\bibfield  {journal} {\bibinfo  {journal} {Phys.
  Rev. B}\ }\textbf {\bibinfo {volume} {92}},\ \bibinfo {pages} {205406}
  (\bibinfo {year} {2015})}\BibitemShut {NoStop}%
\bibitem [{\citenamefont {Gonz\'alez-Tudela}\ and\ \citenamefont
  {Cirac}(2018)}]{Tudela2018}%
  \BibitemOpen
  \bibfield  {author} {\bibinfo {author} {\bibfnamefont {A.}~\bibnamefont
  {Gonz\'alez-Tudela}}\ and\ \bibinfo {author} {\bibfnamefont {J.~I.}\
  \bibnamefont {Cirac}},\ }\bibfield  {title} {\enquote {\bibinfo {title}
  {Exotic quantum dynamics and purely long-range coherent interactions in dirac
  conelike baths},}\ }\href {\doibase 10.1103/PhysRevA.97.043831} {\bibfield
  {journal} {\bibinfo  {journal} {Phys. Rev. A}\ }\textbf {\bibinfo {volume}
  {97}},\ \bibinfo {pages} {043831} (\bibinfo {year} {2018})}\BibitemShut
  {NoStop}%
\bibitem [{\citenamefont {Witzens}, \citenamefont {Loncar},\ and\ \citenamefont
  {Scherer}(2002)}]{Witzens2002}%
  \BibitemOpen
  \bibfield  {author} {\bibinfo {author} {\bibfnamefont {J.}~\bibnamefont
  {Witzens}}, \bibinfo {author} {\bibfnamefont {M.}~\bibnamefont {Loncar}}, \
  and\ \bibinfo {author} {\bibfnamefont {A.}~\bibnamefont {Scherer}},\
  }\bibfield  {title} {\enquote {\bibinfo {title} {Self-collimation in planar
  photonic crystals},}\ }\href {\doibase 10.1109/JSTQE.2002.806693} {\bibfield
  {journal} {\bibinfo  {journal} {IEEE Journal of Selected Topics in Quantum
  Electronics}\ }\textbf {\bibinfo {volume} {8}},\ \bibinfo {pages}
  {1246--1257} (\bibinfo {year} {2002})}\BibitemShut {NoStop}%
\bibitem [{\citenamefont {Mekis}\ \emph {et~al.}(1999)\citenamefont {Mekis},
  \citenamefont {Meier}, \citenamefont {Dodabalapur}, \citenamefont {Slusher},\
  and\ \citenamefont {Joannopoulos}}]{Mekis1999}%
  \BibitemOpen
  \bibfield  {author} {\bibinfo {author} {\bibfnamefont {A.}~\bibnamefont
  {Mekis}}, \bibinfo {author} {\bibfnamefont {M.}~\bibnamefont {Meier}},
  \bibinfo {author} {\bibfnamefont {A.}~\bibnamefont {Dodabalapur}}, \bibinfo
  {author} {\bibfnamefont {R.}~\bibnamefont {Slusher}}, \ and\ \bibinfo
  {author} {\bibfnamefont {J.}~\bibnamefont {Joannopoulos}},\ }\bibfield
  {title} {\enquote {\bibinfo {title} {Lasing mechanism in two-dimensional
  photonic crystal lasers},}\ }\href@noop {} {\bibfield  {journal} {\bibinfo
  {journal} {Applied Physics A}\ }\textbf {\bibinfo {volume} {69}},\ \bibinfo
  {pages} {111--114} (\bibinfo {year} {1999})}\BibitemShut {NoStop}%
\bibitem [{\citenamefont {Galve}\ \emph {et~al.}(2017)\citenamefont {Galve},
  \citenamefont {Mandarino}, \citenamefont {Paris}, \citenamefont {Benedetti},\
  and\ \citenamefont {Zambrini}}]{Galve2017}%
  \BibitemOpen
  \bibfield  {author} {\bibinfo {author} {\bibfnamefont {F.}~\bibnamefont
  {Galve}}, \bibinfo {author} {\bibfnamefont {A.}~\bibnamefont {Mandarino}},
  \bibinfo {author} {\bibfnamefont {M.}~\bibnamefont {Paris}}, \bibinfo
  {author} {\bibfnamefont {C.}~\bibnamefont {Benedetti}}, \ and\ \bibinfo
  {author} {\bibfnamefont {R.}~\bibnamefont {Zambrini}},\ }\bibfield  {title}
  {\enquote {\bibinfo {title} {Microscopic description for the emergence of
  collective dissipation in extended quantum systems},}\ }\href@noop {}
  {\bibfield  {journal} {\bibinfo  {journal} {Scientific Reports}\ }\textbf
  {\bibinfo {volume} {7}},\ \bibinfo {pages} {42050} (\bibinfo {year}
  {2017})}\BibitemShut {NoStop}%
\bibitem [{\citenamefont {John}\ and\ \citenamefont {Wang}(1990)}]{John1990}%
  \BibitemOpen
  \bibfield  {author} {\bibinfo {author} {\bibfnamefont {S.}~\bibnamefont
  {John}}\ and\ \bibinfo {author} {\bibfnamefont {J.}~\bibnamefont {Wang}},\
  }\bibfield  {title} {\enquote {\bibinfo {title} {Quantum electrodynamics near
  a photonic band gap: Photon bound states and dressed atoms},}\ }\href
  {\doibase 10.1103/PhysRevLett.64.2418} {\bibfield  {journal} {\bibinfo
  {journal} {Phys. Rev. Lett.}\ }\textbf {\bibinfo {volume} {64}},\ \bibinfo
  {pages} {2418--2421} (\bibinfo {year} {1990})}\BibitemShut {NoStop}%
\bibitem [{\citenamefont {Kurizki}(1990)}]{Kurizki1990}%
  \BibitemOpen
  \bibfield  {author} {\bibinfo {author} {\bibfnamefont {G.}~\bibnamefont
  {Kurizki}},\ }\bibfield  {title} {\enquote {\bibinfo {title} {Two-atom
  resonant radiative coupling in photonic band structures},}\ }\href {\doibase
  10.1103/PhysRevA.42.2915} {\bibfield  {journal} {\bibinfo  {journal} {Phys.
  Rev. A}\ }\textbf {\bibinfo {volume} {42}},\ \bibinfo {pages} {2915--2924}
  (\bibinfo {year} {1990})}\BibitemShut {NoStop}%
\bibitem [{\citenamefont {Douglas}\ \emph {et~al.}(2015)\citenamefont
  {Douglas}, \citenamefont {Habibian}, \citenamefont {Hung}, \citenamefont
  {Gorshkov}, \citenamefont {Kimble},\ and\ \citenamefont
  {Chang}}]{Douglas2015}%
  \BibitemOpen
  \bibfield  {author} {\bibinfo {author} {\bibfnamefont {J.~S.}\ \bibnamefont
  {Douglas}}, \bibinfo {author} {\bibfnamefont {H.}~\bibnamefont {Habibian}},
  \bibinfo {author} {\bibfnamefont {C.-L.}\ \bibnamefont {Hung}}, \bibinfo
  {author} {\bibfnamefont {A.~V.}\ \bibnamefont {Gorshkov}}, \bibinfo {author}
  {\bibfnamefont {H.~J.}\ \bibnamefont {Kimble}}, \ and\ \bibinfo {author}
  {\bibfnamefont {D.~E.}\ \bibnamefont {Chang}},\ }\bibfield  {title} {\enquote
  {\bibinfo {title} {Quantum many-body models with cold atoms coupled to
  photonic crystals},}\ }\href {\doibase 10.1038/NPHOTON.2015.57} {\bibfield
  {journal} {\bibinfo  {journal} {Nature Photonics}\ }\textbf {\bibinfo
  {volume} {9}},\ \bibinfo {pages} {326--331} (\bibinfo {year}
  {2015})}\BibitemShut {NoStop}%
\bibitem [{\citenamefont {Eriksen}, \citenamefont {Daria},\ and\ \citenamefont
  {Gluckstad}(2002)}]{Eriksen2002}%
  \BibitemOpen
  \bibfield  {author} {\bibinfo {author} {\bibfnamefont {R.~L.}\ \bibnamefont
  {Eriksen}}, \bibinfo {author} {\bibfnamefont {V.~R.}\ \bibnamefont {Daria}},
  \ and\ \bibinfo {author} {\bibfnamefont {J.}~\bibnamefont {Gluckstad}},\
  }\bibfield  {title} {\enquote {\bibinfo {title} {Fully dynamic multiple-beam
  optical tweezers},}\ }\href@noop {} {\bibfield  {journal} {\bibinfo
  {journal} {Optics Express}\ }\textbf {\bibinfo {volume} {10}},\ \bibinfo
  {pages} {597--602} (\bibinfo {year} {2002})}\BibitemShut {NoStop}%
\bibitem [{\citenamefont {Wen}\ \emph {et~al.}(2008)\citenamefont {Wen},
  \citenamefont {David}, \citenamefont {Checoury}, \citenamefont {Kurdi},\ and\
  \citenamefont {Boucaud}}]{Wen2008}%
  \BibitemOpen
  \bibfield  {author} {\bibinfo {author} {\bibfnamefont {F.}~\bibnamefont
  {Wen}}, \bibinfo {author} {\bibfnamefont {S.}~\bibnamefont {David}}, \bibinfo
  {author} {\bibfnamefont {X.}~\bibnamefont {Checoury}}, \bibinfo {author}
  {\bibfnamefont {M.~E.}\ \bibnamefont {Kurdi}}, \ and\ \bibinfo {author}
  {\bibfnamefont {P.}~\bibnamefont {Boucaud}},\ }\bibfield  {title} {\enquote
  {\bibinfo {title} {Two-dimensional photonic crystals with large complete
  photonic band gaps in both te and tm polarizations},}\ }\href {\doibase
  10.1364/OE.16.012278} {\bibfield  {journal} {\bibinfo  {journal} {Optics
  Express}\ }\textbf {\bibinfo {volume} {16}},\ \bibinfo {pages} {12278--12289}
  (\bibinfo {year} {2008})}\BibitemShut {NoStop}%
\bibitem [{\citenamefont {Lodahl}\ \emph {et~al.}(2017)\citenamefont {Lodahl},
  \citenamefont {Mahmoodian}, \citenamefont {Stobbe}, \citenamefont
  {Rauschenbeutel}, \citenamefont {Schneeweiss}, \citenamefont {Volz},
  \citenamefont {Pichler},\ and\ \citenamefont {Zoller}}]{Lodahl2017}%
  \BibitemOpen
  \bibfield  {author} {\bibinfo {author} {\bibfnamefont {P.}~\bibnamefont
  {Lodahl}}, \bibinfo {author} {\bibfnamefont {S.}~\bibnamefont {Mahmoodian}},
  \bibinfo {author} {\bibfnamefont {S.}~\bibnamefont {Stobbe}}, \bibinfo
  {author} {\bibfnamefont {A.}~\bibnamefont {Rauschenbeutel}}, \bibinfo
  {author} {\bibfnamefont {P.}~\bibnamefont {Schneeweiss}}, \bibinfo {author}
  {\bibfnamefont {J.}~\bibnamefont {Volz}}, \bibinfo {author} {\bibfnamefont
  {H.}~\bibnamefont {Pichler}}, \ and\ \bibinfo {author} {\bibfnamefont
  {P.}~\bibnamefont {Zoller}},\ }\bibfield  {title} {\enquote {\bibinfo {title}
  {Chiral quantum optics},}\ }\href {\doibase 10.1038/nature21037} {\bibfield
  {journal} {\bibinfo  {journal} {Nature}\ }\textbf {\bibinfo {volume} {541}},\
  \bibinfo {pages} {473–480} (\bibinfo {year} {2017})}\BibitemShut {NoStop}%
\bibitem [{\citenamefont {Schrader}\ \emph {et~al.}(2001)\citenamefont
  {Schrader}, \citenamefont {Kuhr}, \citenamefont {Alt}, \citenamefont
  {Muller}, \citenamefont {Gomer},\ and\ \citenamefont
  {Meschede}}]{Schrader2001}%
  \BibitemOpen
  \bibfield  {author} {\bibinfo {author} {\bibfnamefont {D.}~\bibnamefont
  {Schrader}}, \bibinfo {author} {\bibfnamefont {S.}~\bibnamefont {Kuhr}},
  \bibinfo {author} {\bibfnamefont {W.}~\bibnamefont {Alt}}, \bibinfo {author}
  {\bibfnamefont {M.}~\bibnamefont {Muller}}, \bibinfo {author} {\bibfnamefont
  {V.}~\bibnamefont {Gomer}}, \ and\ \bibinfo {author} {\bibfnamefont
  {D.}~\bibnamefont {Meschede}},\ }\bibfield  {title} {\enquote {\bibinfo
  {title} {An optical conveyor belt for single neutral atoms},}\ }\href
  {\doibase 10.1007/s003400100722} {\bibfield  {journal} {\bibinfo  {journal}
  {Applied Physics B}\ }\textbf {\bibinfo {volume} {73}},\ \bibinfo {pages}
  {819--824} (\bibinfo {year} {2001})}\BibitemShut {NoStop}%
\bibitem [{\citenamefont {Burgers}\ \emph {et~al.}(2018)\citenamefont
  {Burgers}, \citenamefont {Peng}, \citenamefont {Muniz}, \citenamefont
  {McClung}, \citenamefont {Martin},\ and\ \citenamefont
  {Kimble}}]{Burgers2018}%
  \BibitemOpen
  \bibfield  {author} {\bibinfo {author} {\bibfnamefont {A.}~\bibnamefont
  {Burgers}}, \bibinfo {author} {\bibfnamefont {L.}~\bibnamefont {Peng}},
  \bibinfo {author} {\bibfnamefont {J.}~\bibnamefont {Muniz}}, \bibinfo
  {author} {\bibfnamefont {A.}~\bibnamefont {McClung}}, \bibinfo {author}
  {\bibfnamefont {M.}~\bibnamefont {Martin}}, \ and\ \bibinfo {author}
  {\bibfnamefont {H.}~\bibnamefont {Kimble}},\ }\bibfield  {title} {\enquote
  {\bibinfo {title} {Clocked atom delivery to a photonic crystal waveguide},}\
  }\href@noop {} {\bibfield  {journal} {\bibinfo  {journal} {arXiv preprint
  arXiv:1810.07757}\ } (\bibinfo {year} {2018})}\BibitemShut {NoStop}%
\bibitem [{\citenamefont {Buhmann}\ and\ \citenamefont
  {Welsch}(2007)}]{Buhmann2007}%
  \BibitemOpen
  \bibfield  {author} {\bibinfo {author} {\bibfnamefont {S.~Y.}\ \bibnamefont
  {Buhmann}}\ and\ \bibinfo {author} {\bibfnamefont {D.-G.}\ \bibnamefont
  {Welsch}},\ }\bibfield  {title} {\enquote {\bibinfo {title} {Dispersion
  forces in macroscopic quantum electrodynamics},}\ }\href {\doibase
  10.1016/j.pquantelec.2007.03.001} {\bibfield  {journal} {\bibinfo  {journal}
  {Progress in Quantum Electronics}\ }\textbf {\bibinfo {volume} {31}},\
  \bibinfo {pages} {51--130} (\bibinfo {year} {2007})}\BibitemShut {NoStop}%
\bibitem [{\citenamefont {Lester}\ \emph {et~al.}(2015)\citenamefont {Lester},
  \citenamefont {Luick}, \citenamefont {Kaufman}, \citenamefont {Reynolds},\
  and\ \citenamefont {Regal}}]{Lester2015}%
  \BibitemOpen
  \bibfield  {author} {\bibinfo {author} {\bibfnamefont {B.~J.}\ \bibnamefont
  {Lester}}, \bibinfo {author} {\bibfnamefont {N.}~\bibnamefont {Luick}},
  \bibinfo {author} {\bibfnamefont {A.~M.}\ \bibnamefont {Kaufman}}, \bibinfo
  {author} {\bibfnamefont {C.~M.}\ \bibnamefont {Reynolds}}, \ and\ \bibinfo
  {author} {\bibfnamefont {C.~A.}\ \bibnamefont {Regal}},\ }\bibfield  {title}
  {\enquote {\bibinfo {title} {Rapid production of uniformly filled arrays of
  neutral atoms},}\ }\href {\doibase 10.1103/PhysRevLett.115.073003} {\bibfield
   {journal} {\bibinfo  {journal} {Physical Review Letters}\ }\textbf {\bibinfo
  {volume} {115}},\ \bibinfo {pages} {073003} (\bibinfo {year}
  {2015})}\BibitemShut {NoStop}%
\bibitem [{\citenamefont {Endres}\ \emph {et~al.}(2016)\citenamefont {Endres},
  \citenamefont {Bernien}, \citenamefont {Keesling}, \citenamefont {Levine},
  \citenamefont {Anschuetz}, \citenamefont {Krajenbrink}, \citenamefont
  {Senko}, \citenamefont {Vuletic}, \citenamefont {Greiner},\ and\
  \citenamefont {Lukin}}]{Endres2016}%
  \BibitemOpen
  \bibfield  {author} {\bibinfo {author} {\bibfnamefont {M.}~\bibnamefont
  {Endres}}, \bibinfo {author} {\bibfnamefont {H.}~\bibnamefont {Bernien}},
  \bibinfo {author} {\bibfnamefont {A.}~\bibnamefont {Keesling}}, \bibinfo
  {author} {\bibfnamefont {H.}~\bibnamefont {Levine}}, \bibinfo {author}
  {\bibfnamefont {E.~R.}\ \bibnamefont {Anschuetz}}, \bibinfo {author}
  {\bibfnamefont {A.}~\bibnamefont {Krajenbrink}}, \bibinfo {author}
  {\bibfnamefont {C.}~\bibnamefont {Senko}}, \bibinfo {author} {\bibfnamefont
  {V.}~\bibnamefont {Vuletic}}, \bibinfo {author} {\bibfnamefont
  {M.}~\bibnamefont {Greiner}}, \ and\ \bibinfo {author} {\bibfnamefont
  {M.~D.}\ \bibnamefont {Lukin}},\ }\bibfield  {title} {\enquote {\bibinfo
  {title} {Atom-by-atom assembly of defect-free one-dimensional cold atom
  arrays},}\ }\href {\doibase 10.1126/science.aah3752} {\bibfield  {journal}
  {\bibinfo  {journal} {Science}\ } (\bibinfo {year} {2016}),\
  10.1126/science.aah3752}\BibitemShut {NoStop}%
\bibitem [{\citenamefont {Barredo}\ \emph {et~al.}(2016)\citenamefont
  {Barredo}, \citenamefont {de~L{\'e}s{\'e}leuc}, \citenamefont {Lienhard},
  \citenamefont {Lahaye},\ and\ \citenamefont {Browaeys}}]{Barredo2016}%
  \BibitemOpen
  \bibfield  {author} {\bibinfo {author} {\bibfnamefont {D.}~\bibnamefont
  {Barredo}}, \bibinfo {author} {\bibfnamefont {S.}~\bibnamefont
  {de~L{\'e}s{\'e}leuc}}, \bibinfo {author} {\bibfnamefont {V.}~\bibnamefont
  {Lienhard}}, \bibinfo {author} {\bibfnamefont {T.}~\bibnamefont {Lahaye}}, \
  and\ \bibinfo {author} {\bibfnamefont {A.}~\bibnamefont {Browaeys}},\
  }\bibfield  {title} {\enquote {\bibinfo {title} {An atom-by-atom assembler of
  defect-free arbitrary two-dimensional atomic arrays},}\ }\href {\doibase
  10.1126/science.aah3778} {\bibfield  {journal} {\bibinfo  {journal}
  {Science}\ }\textbf {\bibinfo {volume} {354}},\ \bibinfo {pages} {1021--1023}
  (\bibinfo {year} {2016})}\BibitemShut {NoStop}%
\bibitem [{\citenamefont {Kim}\ \emph {et~al.}(2018)\citenamefont {Kim},
  \citenamefont {Chang}, \citenamefont {Fields}, \citenamefont {Chen},\ and\
  \citenamefont {Hung}}]{Kim2018}%
  \BibitemOpen
  \bibfield  {author} {\bibinfo {author} {\bibfnamefont {M.~E.}\ \bibnamefont
  {Kim}}, \bibinfo {author} {\bibfnamefont {T.-H.}\ \bibnamefont {Chang}},
  \bibinfo {author} {\bibfnamefont {B.~M.}\ \bibnamefont {Fields}}, \bibinfo
  {author} {\bibfnamefont {C.-A.}\ \bibnamefont {Chen}}, \ and\ \bibinfo
  {author} {\bibfnamefont {C.-L.}\ \bibnamefont {Hung}},\ }\bibfield  {title}
  {\enquote {\bibinfo {title} {Trapping single atoms on a nanophotonic circuit
  with configurable tweezer lattices},}\ }\href@noop {} {\bibfield  {journal}
  {\bibinfo  {journal} {arXiv preprint arXiv:1810.08769}\ } (\bibinfo {year}
  {2018})}\BibitemShut {NoStop}%
\bibitem [{\citenamefont {Cohen}, \citenamefont {Meenehan},\ and\ \citenamefont
  {Painter}(2013)}]{Cohen2013}%
  \BibitemOpen
  \bibfield  {author} {\bibinfo {author} {\bibfnamefont {J.~D.}\ \bibnamefont
  {Cohen}}, \bibinfo {author} {\bibfnamefont {S.~M.}\ \bibnamefont {Meenehan}},
  \ and\ \bibinfo {author} {\bibfnamefont {O.}~\bibnamefont {Painter}},\
  }\bibfield  {title} {\enquote {\bibinfo {title} {Optical coupling to
  nanoscale optomechanical cavities for near quantum-limited motion
  transduction},}\ }\href {\doibase 10.1364/OE.21.011227} {\bibfield  {journal}
  {\bibinfo  {journal} {Optics Express}\ }\textbf {\bibinfo {volume} {21}},\
  \bibinfo {pages} {11227--11236} (\bibinfo {year} {2013})}\BibitemShut
  {NoStop}%
\bibitem [{\citenamefont {Lee}\ \emph {et~al.}(2008)\citenamefont {Lee},
  \citenamefont {Choi}, \citenamefont {Kim}, \citenamefont {Park},\ and\
  \citenamefont {Kee}}]{Lee08}%
  \BibitemOpen
  \bibfield  {author} {\bibinfo {author} {\bibfnamefont {S.-G.}\ \bibnamefont
  {Lee}}, \bibinfo {author} {\bibfnamefont {J.-S.}\ \bibnamefont {Choi}},
  \bibinfo {author} {\bibfnamefont {J.-E.}\ \bibnamefont {Kim}}, \bibinfo
  {author} {\bibfnamefont {H.~Y.}\ \bibnamefont {Park}}, \ and\ \bibinfo
  {author} {\bibfnamefont {C.-S.}\ \bibnamefont {Kee}},\ }\bibfield  {title}
  {\enquote {\bibinfo {title} {Reflection minimization at two-dimensional
  photonic crystal interfaces},}\ }\href {\doibase 10.1364/OE.16.004270}
  {\bibfield  {journal} {\bibinfo  {journal} {Opt. Express}\ }\textbf {\bibinfo
  {volume} {16}},\ \bibinfo {pages} {4270--4277} (\bibinfo {year}
  {2008})}\BibitemShut {NoStop}%
\bibitem [{\citenamefont {Tanaka}\ \emph {et~al.}(2004)\citenamefont {Tanaka},
  \citenamefont {Sugimoto}, \citenamefont {Ikeda}, \citenamefont {Nakamura},
  \citenamefont {Asakawa}, \citenamefont {Inoue},\ and\ \citenamefont
  {Johnson}}]{Tanaka2004}%
  \BibitemOpen
  \bibfield  {author} {\bibinfo {author} {\bibfnamefont {Y.}~\bibnamefont
  {Tanaka}}, \bibinfo {author} {\bibfnamefont {Y.}~\bibnamefont {Sugimoto}},
  \bibinfo {author} {\bibfnamefont {N.}~\bibnamefont {Ikeda}}, \bibinfo
  {author} {\bibfnamefont {H.}~\bibnamefont {Nakamura}}, \bibinfo {author}
  {\bibfnamefont {K.}~\bibnamefont {Asakawa}}, \bibinfo {author} {\bibfnamefont
  {K.}~\bibnamefont {Inoue}}, \ and\ \bibinfo {author} {\bibfnamefont
  {S.}~\bibnamefont {Johnson}},\ }\bibfield  {title} {\enquote {\bibinfo
  {title} {Group velocity dependence of propagation losses in
  single-line-defect photonic crystal waveguides on gaas membranes},}\ }\href
  {\doibase 10.1049/el:20040114} {\bibfield  {journal} {\bibinfo  {journal}
  {Electronics Letters}\ }\textbf {\bibinfo {volume} {40}},\ \bibinfo {pages}
  {174--176} (\bibinfo {year} {2004})}\BibitemShut {NoStop}%
\bibitem [{\citenamefont {Hung}\ \emph {et~al.}(2016)\citenamefont {Hung},
  \citenamefont {Gonz\'alez-Tudela}, \citenamefont {Cirac},\ and\ \citenamefont
  {Kimble}}]{Hung2016}%
  \BibitemOpen
  \bibfield  {author} {\bibinfo {author} {\bibfnamefont {C.-L.}\ \bibnamefont
  {Hung}}, \bibinfo {author} {\bibfnamefont {A.}~\bibnamefont
  {Gonz\'alez-Tudela}}, \bibinfo {author} {\bibfnamefont {J.~I.}\ \bibnamefont
  {Cirac}}, \ and\ \bibinfo {author} {\bibfnamefont {H.~J.}\ \bibnamefont
  {Kimble}},\ }\bibfield  {title} {\enquote {\bibinfo {title} {Quantum spin
  dynamics with pairwise-tunable, long-range interactions},}\ }\href {\doibase
  10.1073/pnas.1603777113} {\bibfield  {journal} {\bibinfo  {journal}
  {Proceedings to National Academy of Science}\ }\textbf {\bibinfo {volume}
  {133}},\ \bibinfo {pages} {E4946--E4955} (\bibinfo {year}
  {2016})}\BibitemShut {NoStop}%
\bibitem [{\citenamefont {{COMSOL AB, Stockholm, Sweden}}(2018)}]{Comsol}%
  \BibitemOpen
  \bibfield  {author} {\bibinfo {author} {\bibnamefont {{COMSOL AB, Stockholm,
  Sweden}}},\ }\href@noop {} {\enquote {\bibinfo {title} {Comsol
  multiphysics®},}\ }\bibinfo {howpublished} {\url{www.comsol.com}} (\bibinfo
  {year} {2018})\BibitemShut {NoStop}%
\bibitem [{\citenamefont {Oskooi}\ \emph {et~al.}(2010)\citenamefont {Oskooi},
  \citenamefont {Roundy}, \citenamefont {Ibanescu}, \citenamefont {Bermel},
  \citenamefont {Joannopoulos},\ and\ \citenamefont {Johnson}}]{Oskoi2010}%
  \BibitemOpen
  \bibfield  {author} {\bibinfo {author} {\bibfnamefont {A.~F.}\ \bibnamefont
  {Oskooi}}, \bibinfo {author} {\bibfnamefont {D.}~\bibnamefont {Roundy}},
  \bibinfo {author} {\bibfnamefont {M.}~\bibnamefont {Ibanescu}}, \bibinfo
  {author} {\bibfnamefont {P.}~\bibnamefont {Bermel}}, \bibinfo {author}
  {\bibfnamefont {J.}~\bibnamefont {Joannopoulos}}, \ and\ \bibinfo {author}
  {\bibfnamefont {S.~G.}\ \bibnamefont {Johnson}},\ }\bibfield  {title}
  {\enquote {\bibinfo {title} {Meep: A flexible free-software package for
  electromagnetic simulations by the fdtd method},}\ }\href {\doibase
  https://doi.org/10.1016/j.cpc.2009.11.008} {\bibfield  {journal} {\bibinfo
  {journal} {Computer Physics Communications}\ }\textbf {\bibinfo {volume}
  {181}},\ \bibinfo {pages} {687 -- 702} (\bibinfo {year} {2010})}\BibitemShut
  {NoStop}%
\bibitem [{\citenamefont {{Lumerical Inc}}(2018)}]{Lumerical}%
  \BibitemOpen
  \bibfield  {author} {\bibinfo {author} {\bibnamefont {{Lumerical Inc}}},\
  }\href@noop {} {}\bibinfo {howpublished}
  {\url{http://www.lumerical.com/tcad-products/fdtd/}} (\bibinfo {year}
  {2018})\BibitemShut {NoStop}%
\end{thebibliography}%

\end{document}